# Magnetic Field-Induced Polar Order in Monolayer Molybdenum Disulfide Transistors

*Duxing Hao, Wen-Hao Chang, Yu-Chen Chang, Wei-Tung Liu, Sheng-Zhu Ho, Chen-Hsuan Lu, Tilo H. Yang, Naoya Kawakami, Yi-Chun Chen, Ming-Hao Liu, Chun-Liang Lin,* Ting-Hua Lu,* Yann-Wen Lan,* and Nai-Chang Yeh**

Duxing Hao, Prof. Nai-Chang Yeh
Department of Physics
California Institute of Technology
Pasadena, California, 91125, USA
E-mail: ncyeh@caltech.edu

Wen-Hao Chang, Yu-Chen Chang, Tilo H. Yang, Prof. Ting-Hua Lu, Prof. Yann-Wen Lan, Prof. Nai-Chang Yeh
Department of Physics
National Taiwan Normal University
Taipei, 116059, Taiwan
E-mail: thlu@ntnu.edu.tw, ywlan@ntnu.edu.tw, ncyeh@caltech.edu

Wei-Tung Liu, Naoya Kawakami, Prof. Chun-Liang Lin
Department of Electrophysics
National Yang Ming Chiao Tung University
Hsinchu, 300093, Taiwan
E-mail: clin@nycu.edu.tw

Sheng-Zhu Ho, Prof. Yi-Chun Chen, Prof. Ming-Hao Liu
Department of Physics
National Cheng Kung University
Tainan, 701401, Taiwan

Chen-Hsuan Lu
Department of Applied Physics and Materials Science
California Institute of Technology
Pasadena, California, 91125, USA

Tilo H. Yang

Department of Electrical Engineering and Computer Science

Massachusetts Institute of Technology

Cambridge, Massachusetts, 02139, USA

Prof. Nai-Chang Yeh

Kavli Nanoscience Institute

California Institute of Technology

Pasadena, California, 91125, USA

E-mail: ncyeh@caltech.edu

In semiconducting monolayer transition metal dichalcogenides (ML-TMDs), broken inversion symmetry and strong spin-orbit coupling result in spin-valley lock-in effects so that the valley degeneracy may be lifted by external magnetic fields, potentially leading to real-space structural transformation. Here, we report magnetic field ($B$)-induced giant electric hysteretic responses to back-gate voltages in ML-$MoS_2$ field-effect transistors (FETs) on $SiO_2$/Si at temperatures < 20 K. The observed hysteresis increases with $|B|$ up to 12 T and is tunable by varying the temperature. Raman spectroscopic and scanning tunneling microscopic studies reveal significant lattice expansion with increasing $|B|$ at 4.2 K, and this lattice expansion becomes asymmetric in ML-$MoS_2$ FETs on rigid $SiO_2$/Si substrates, leading to out-of-plane mirror symmetry breaking and the emergence of a tunable out-of-plane ferroelectric-like polar order. This broken symmetry-induced polarization in ML-$MoS_2$ shows typical ferroelectric butterfly hysteresis in piezo-response force microscopy, adding ML-$MoS_2$ to the single-layer material family that exhibit out-of-plane polar order-induced ferroelectricity, which is promising for such technological applications as cryo-temperature ultracompact non-volatile memories, memtransistors, and ultrasensitive magnetic field sensors. Moreover, the polar effect induced by asymmetric lattice expansion may be further generalized to other ML-TMDs and achieved by nanoscale strain engineering of the substrate without magnetic fields.

# 1. Introduction

Polar order and its associated hysteretic behavior have played critically important roles in modern electronics applications. Ferroelectric materials, which feature a polar point group and polarization hysteresis upon switching external electric field, are crucial for high-density data storage, microwave devices, pyroelectric sensors, and non-volatile memories.[1–7] With the rapidly increasing demand for data storage capabilities, there is strong desire to reduce the dimensions of ferroelectric devices by exploring new materials, including 2D semiconductors such as 2H-TMDs. However, in the case of monolayer (ML) 2H-TMDs, the $D_{3h}$ point group (space group P6m2) preserves the centrosymmetry and is therefore incompatible with ferroelectricity. If mirror symmetry about the horizontal plane of transition metal layer can be broken, such as in the case of ML-$MoS_2$ with a distorted-1T structure ($C_{3v}$, P3m1) where a slight vertical displacement of intralayer sulfur atoms results in broken mirror symmetry and a net electric polarization along the out-of-plane direction, then the lattice space group may reduce into a ferroelectric subgroup[8] and thus the emergence of ferroelectricity.[9–11] In this context, it is interesting to note that the shear-transformed bilayer 3R-$MoS_2$ system with broken mirror symmetry between the two $MoS_2$ monolayers reduces the symmetry group to $C_{3v}$ (R3m), where ferroelectric response has been discovered recently.[12] Similarly, ferroelectricity has been reported in van der Waals (vdW) interfaces between two marginally twisted ML-vdW materials that lack polar point groups in their parent lattices, such as boron nitride (BN) and transition metal dichalcogenides (TMDs).[13–15]

Here we report giant ferroelectric-like hysteresis induced by out-of-plane magnetic field ($B$) in field-effect transistors (FETs) based on ML-$MoS_2$ single crystals on $SiO_2$/Si substrates at temperatures ($T$) below ~20 K. The counterclockwise hysteresis of the source-drain current ($I_{DS}$) as a function of the back-gate voltage ($V_{GS}$) can be enhanced by increasing the maximum $V_{GS}$ value, increasing $|B|$, and lowering $T$. These robust findings from five distinct ML-$MoS_2$ FETs at low temperatures and under finite magnetic fields differ drastically from previous reports of high-temperature and zero-$B$ hysteretic behavior in ML-$MoS_2$ FETs, the latter were attributed to mechanisms such as thermally-activated trapped states[16–18] absorbates,[19,20] and ordinary gate voltage-induced stress effects,[21] where clockwise hysteresis loops were observed near room temperatures without magnetic field dependences. We further present experimental evidence for magnetic field-induced lattice expansion in ML-$MoS_2$, which is associated with real-space structural transformation from lifting the valley degeneracy by magnetic field through spin-valley lock-in in ML-$MoS_2$. Based on these observations, we attribute the appearance of ferroelectric-like polar effects to asymmetric lattice expansion of ML-MoS$_2$ on rigid SiO$_2$/Si substrates, and we further corroborate this conjecture by demonstrating absence of polar effects in FETs consisting of ML-MoS$_2$ on substrates buffered with multilayers of hexagonal boron nitride (h-BN) with a thermal expansion coefficient (TEC) comparable to that of MoS$_2$ to prevent the occurrence of asymmetric lattice expansion.

The MoS$_2$-FETs studied in this work were based on high quality 1H-MoS$_2$ single crystals (*a.k.a.* ML-MoS$_2$ with the 2H-phase) grown by chemical vapor deposition (CVD).[22–24] The schematic of a transistor and its electrical measurement circuit is shown in the inset of **Figure 1**a, where the source-drain voltage ($V_{DS}$) was applied between a pair of bismuth (Bi)/gold (Au) contacts, and the gate voltage ($V_{GS}$) was applied between the source contact and a heavily *p*-doped Si substrate with a 30 nm-thick SiO$_2$ insulating layer. The ML-MoS$_2$ single crystals exhibited high degrees of homogeneity and were free from magnetic impurities, as verified by their optical and x-ray photoelectron spectroscopic characterizations exemplified in Methods and Figure S1-3. Nearly ohmic source-drain current ($I_{DS}$) versus gate voltage ($V_{GS}$) transfer curves were observed down to 1.8 K (Figure 1a) due to ultralow-Schottky barriers associated with the Bi/Au contacts[25] (Figure S4, Figure S5). The mobility of these FET devices was typically around ~ 10 cm$^2$V$^{-1}$S$^{-1}$ or better at 300 K, as detailed previously.[26] These values was comparable to the reported mobility values ranging 0.1-10 cm$^2$V$^{-1}$S$^{-1}$ for typical FET devices made of exfoliated ML-MoS$_2$ on SiO$_2$-supported substrates,[27–29] which appears to be limited by substrate charged impurities instead of sample quality-limited or contact-limited.[30,31]

## 2. Results and Discussion

### 2.1. Emerging electric hysteresis in MoS$_2$-FETs under magnetic field

Upon application of an out-of-plane magnetic field ($B$), counterclockwise $|I_{DS}|$-*vs.*-$V_{GS}$ hysteretic loops emerged from measurements of the FET devices at 1.8 K, as exemplified in Figure 1b for $B = 9$ T, where the $|I_{DS}|$-*vs.*-$V_{GS}$ transfer curves were measured with different $V_{DS}$ values fixed at −1.0 V, −0.8 V, and −0.6 V. On the other hand, no loop was present for $B = 0$ (Figure 1b inset). This emergence of counterclockwise hysteretic behavior at low temperature and finite magnetic field is a response of ferroelectric-like polar order modulation caused by magnetic-field induced asymmetric rippling effect, which is fundamentally different from previously known hysteresis-inducing mechanisms as elaborated in Supplementary Note 1. When sweeping up $V_{GS}$, the system was initially in the high-resistance state (HRS),[32] showing $|I_{DS}| > 0$ for $V_{GS} > V_{th,H}$, where $V_{th,H}$ denoted the threshold voltage for the forward branch defined in Methods. In contrast, when $V_{GS}$ was reduced from a finite $|I_{DS}|$ state, the lattice returned from a highly polarized low-resistance state (LRS)[1,32] so that $|I_{DS}|$ remained finite until $V_{GS}$ reached

$V_{th,L}$ ($< V_{th,H}$), where $V_{th,L}$ represented the threshold voltage for $|I_{DS}| > 0$ in the returned branch. Within the applicable range of $V_{GS}$ up to 36 V in all 5 devices (Dev. #1 - #5), none of the hysteresis loops became fully closed within our experimental parameters due to limited electrical doping capability and the necessity of keeping the leakage current small. We found that $|I_{DS}|$ continued to increase upon reversing $V_{GS}$ from 20 V to 16 V (Figure 1b), which may be due to the transient response of charging as a result of polar order modulation. Additionally, while the $|I_{DS}|$-$vs.$-$V_{GS}$ curves measured with different $V_{DS}$ values yielded different loop shapes, the extracted hysteresis window $V_{HW}$ ($\equiv V_{th,H} - V_{th,L}$) remained the same (Figure S6 and Figure S7). The independence of $V_{HW}$ on the in-plane voltage $V_{DS}$ suggested that the field-induced polarization was mostly tunable via the out-of-plane electric field. Further studies of the $|I_{DS}|$-$vs.$-$V_{GS}$ curves for different $B$ values under a constant $V_{DS} = -1.0$ V are shown in Figure 1c for one of the devices.

Overall, the following key findings were consistently observed across five different devices (Figure S8): (1) under a constant out-of-plane $B$, the $|I_{DS}|$ value obtained at a given $V_{GS}$ was smaller for a larger $|B|$ due to a positive magnetoresistance (Figure S9); (2) the size of the hysteresis loop $V_{HW}$ increased with $|B|$, suggesting a stronger polar order under a higher magnetic field; (3) the hysteresis loop was independent of the sign of $B$ from $-9$ T to 9 T (Figure 1c); (4) no discernible magneto-hysteresis of $|I_{DS}|$-$vs.$-$V_{GS}$ curves was observed during the two consecutive magnetic field sweeps from $-9$ T to 9 T then from 9 T back to $-9$ T (Figure S10), indicating absence of out-of-plane magnetism; and (5) the hysteresis window $V_{HW}$ increased approximately linearly with $|B|$ for $|B| < 3$ T, and then saturated for large $|B|$ (Figure 1d).

In addition to magnetic field, gate voltage and temperature may be used to modulate this hysteresis, although the resulting effects are less significant than using the magnetic field. **Figure 2**a shows the linear and semi-log scale (inset) $|I_{DS}|$-$vs.$-$V_{GS}$ curves of a FET device measured at 10 K and $B = 9$ T and for $V_{GS}$ swept from 2 V to different maximum gate voltages ($V_{GS, max}$) then back. Interestingly, under a fixed magnetic field, a higher $V_{GS, max}$ led to increasing although eventually saturating $V_{HW}$ (Figure 2b). On the other hand, upon increasing temperature from 1.8 K under a constant magnetic field of 12 T, the size of the hysteresis loop first shrank gradually with temperature, and then decreased precipitously above ~14 K, and finally vanished at the critical temperature ($T_C$) ~19.3 K (Figure 2c) where $V_{HW}$ vanished completely (Figure 2c, inset). Above the $T_C$, a small magnetic field-independent clockwise hysteresis background emerged as exemplified in Figure S11, which was due to the gate voltage stress[21] and remnant oxide trapping close to the $SiO_2$-$MoS_2$ interface as previously reported.[16,17] We further investigated the temperature dependence of $V_{HW}$ for the counterclockwise hysteresis loops under different constant magnetic fields (Figure 2d) and found that the critical temperature $T_C$ where $V_{HW}$ vanished increased with increasing $|B|$ (Figure 2d, inset), whereas for a given temperature below $T_C$, $V_{HW}$ always increased with $|B|$. Noting that the in-plane TA and LA phonon modes of ML-MoS$_2$ became frozen below 20 K[33] where the magnetic field induced lattice expansion emerged, the modulation of $T_C$ by magnetic field was less disturbed by in-plane phonons so that larger degrees of lattice expansion induced by stronger magnetic field may survive a higher temperature, yielding the observation of a higher $T_C$. Overall, comparable $T_C$-vs.-$B$ values within ±1 K variations among four different devices were found (Figure S12), suggesting the robust presence of this hysteresis across devices.

## 2.2. Magnetic field-induced MoS$_2$ lattice expansion

To elucidate the physical origin for magnetic field-induced hysteresis in these ML-MoS$_2$ FETs, we carried out cryo-temperature Raman spectroscopic studies of ML-MoS$_2$ on SiO$_2$/Si substrates, as shown in **Figure 3**a and 3b. Upon cooling the sample from 300 K at $B = 0$, both the A$_{1g}$ and E$_{2g}$ peaks were found blue-shifted due to a positive thermal expansion coefficient (TEC) down to ~20 K, below which the TEC became slightly negative as a result of frozen in-plane phonon modes[33–35] (Figure 3c and Figure S13). In particular, the peak positions of the A$_{1g}$ and E$_{2g}$ modes at 4.2 K were found to be at 406.6 cm$^{-1}$ and 388.6 cm$^{-1}$, respectively, both higher than the A$_{1g}$ (404.3 cm$^{-1}$) and E$_{2g}$ (386.7 cm$^{-1}$) peak positions measured at 300 K. Upon the application of an out-of-plane magnetic field of 0.5 T, however, significant field-induced redshifts were found for both the E$_{2g}$ (−3.1 cm$^{-1}$) and A$_{1g}$ (−1.5 cm$^{-1}$) modes at 4.2 K, whereas no discernible field-induced effect on the Raman modes was observed at 300 K, as shown in Figure 3c. These findings suggest that under a magnetic field below $T_C(B)$, both the in-plane E$_{2g}$ mode and the out-of-plane A$_{1g}$ mode became softened, which implied a tensile strain on the lattice, similar to the Raman mode softening observed in the bubbled regions of ML-WS$_2$ encapsulated by h-BN.[36]

Further evidence for magnetic field-induced lattice expansion was manifested by scanning tunneling microscopic (STM) studies of a ML-MoS$_2$ sample grown *in situ* on highly ordered pyrolytic graphite (HOPG). The filtered ML-MoS$_2$/HOPG topographic images (Figure S14 and Figure S15) showed evolving moiré patterns with magnetic field for two distinct (5 nm × 5 nm) regions with a twist angle ($\varphi$) of 0.5° (**Figure 4**a) and 3.1° (Figure 4b) at $B = 0$, respectively. The corresponding reciprocal lattice points for the moiré pattern and ML-MoS$_2$ were respectively highlighted in red and blue circles in the Fast Fourier transformation (FFT) graphs of the topography, as shown in the bottom panels of Figure 4a and Figure 4b. With increasing $B$, a systematically increasing ML-MoS₂ lattice constant as well as a decreasing moiré lattice constant were found from analyzing the reciprocal lattice vectors in the corresponding FFT graphs taken over the same (5 nm × 5 nm) area, whereas the HOPG lattice constant remained the same.[37] As shown in Figure 4c, we found ~3% in-plane lattice expansion for $B = 5$ T and ~1.3% in-plane lattice expansion for $B = 0.5$ T, the latter agreed well with the estimate of ~1.4% lattice expansion from the redshift of the $E_{2g}$ Raman mode.[38,39] These results thus provide solid evidence for substantial magnetic field-induced lattice expansion in ML-MoS₂ at low temperature. Moreover, the expected $A_{1g}$ shift from the pure tensile strain of 1.4% would have been about −0.8 cm⁻¹ as predicted by first principle calculations and previous experiments,[38,39] yet a larger shift of −1.5cm⁻¹ was observed in our Raman measurement as shown in Figure 3c. This difference may be understood by noting that the peak position of the out-of-plane $A_{1g}$ phonon mode is much more sensitive to charge accumulation than the in-plane $E_{2g}$ mode,[39] so that the larger redshift found in the $A_{1g}$ mode may be attributed to the presence of net out-of-plane charge distributions[40] and thus supports the occurrence of spontaneous out-of-plane polarization.

## 2.3. Asymmetric lattice expansion-induced ferroelectric-like polar order

A plausible explanation for our observation of magnetic field-induced giant hysteretic behavior is due to a bistable ferroelectric-like spontaneous out-of-plane polarization, which emerges under anisotropic magnetic field-induced lattice expansion due to the TEC mismatch between ML-MoS₂ and the underlying rigid SiO₂/Si substrate,[33,41,42] leading to asymmetric flexoelectric effect[41] as schematically shown in Figure 4d and Figure 4e. Further supporting evidence for this scenario was provided by piezo-response force microscopy (PFM) measurements, which revealed evident ferroelectric butterfly hysteresis loop emerged at 1.6 K and $B = 3$ T whereas no piezoelectric response was found at $B = 0$, as exemplified in Figure S16. This finding indicated that the out-of-plane polar order under a finite out-of-plane magnetic field was switchable by a nonlinear electric field applied by a PFM.[41]

Other than net out-of-plane polarization due to asymmetric magnetic field-induced lattice expansion and rippling in ML-MoS₂ on rigid substrate, local charge disorder due to sulfur vacancies could also contribute to the nonlinear effect found in the PFM measurements. However, the tendency of reduced tensile strain loading with increasing vacancy concentration as the result of decreasing Young's modulus[43] would have led to diminished flexoelectric effect with increasing sulfur vacancies. Moreover, the low sulfur vacancy concentrations (~ 0.2%) in all ML-MoS₂ samples (Figure S17 and Figure S18) that exhibited polar effects was consistent with insignificant roles of sulfur vacancies in the magnetic field-induced ferroelectric responses.

The important role of the substrate in the occurrence of out-of-plane asymmetric lattice expansion and the resulting polar effect was further verified by studying two distinct ML-MoS$_2$ FETs buffered by ~5 nm thick h-BN between the ML-MoS$_2$ and the rigid SiO$_2$/Si substrate. Both buffered FETs revealed no obvious magnetic field-induced counterclockwise hysteresis (Figure S16), which was expected because the h-BN buffer had a similar negative TEC[44] and also exhibited low lateral friction[45] relative to the MoS$_2$ layer so that the centrosymmetry-breaking required for ferroelectric-like out-of-plane polarization was largely prevented in the buffered ML-MoS$_2$ FETs on h-BN/SiO$_2$/Si substrates (Figure S19). The relevance of the substrate to the occurrence of out-of-plane polar order in ML-MoS$_2$ is schematically illustrated in Figure 4d, where the side-views for the asymmetric and symmetric rippling effects between the top and bottom sulfur layers are shown for ML-MoS$_2$ on SiO$_2$/Si and h-BN/SiO$_2$/Si substrates, respectively. Our proposed model of asymmetric lattice expansion between the top and bottom sulfur layers of ML-MoS$_2$ on SiO$_2$/Si below $T_C$ and under an out-of-plane external magnetic field is further illustrated in Figure 4e, showing the occurrence of broken mirror symmetry.

## 3. Conclusion

Overall, our fully reproducible experimental findings from five distinct FET devices of ML-MoS$_2$ on SiO$_2$/Si substrates suggested that the out-of-plane magnetic field-induced ferroelectric-like counterclockwise hysteresis was robust at low temperature and reversible upon removing the magnetic field, and the absence of such phenomena in two buffered FET devices of ML-MoS$_2$ on h-BN/SiO$_2$/Si further accentuated the important role of rigid substrates in inducing the asymmetric lattice expansion, which was essential for the out-of-plane electric polarization. The magnetic field-induced lattice expansion in ML-MoS$_2$ at low temperatures is likely associated with the occurrence of a magnetic field-induced structural phase transformation, although the microscopic mechanism and the nature of this phase transformation remain unclear. A possibility may be related to lifting the valley degeneracy in ML-MoS$_2$ by magnetic field via spin-valley coupling, thereby resulting in a real-space structural transformation. Additionally, the correlation between the size of hysteresis ($V_{HW}$) and the magnitude of out-of-plane magnetic field ($|B|$) is suggestive of multiferroic-like behavior. Overall, a careful *ab initio* calculation that takes into consideration of the effects of magnetic field, temperature, sulfur vacancies, and substrate will be necessary to fully account for our observation and to unravel the underlying physical mechanism, which is beyond the current scope of our work. Regardless of the microscopic physical origin, the giant magnetic-field induced ferroelectric-like responses in the ML-MoS$_2$ FET devices exhibited strong stability and reproducibility, thus promising for such technological applications as cryo-temperature ultracompact non-volatile memories, memtransistors, and ultrasensitive magnetic field sensors. Furthermore, the observed ferroelectric behavior induced by asymmetric lattice expansion could potentially be generalized to other monolayer electronics, which opens exciting new possibilities of using nanoscale strain engineering of the substrate to achieve similar effects in various two-dimensional materials, paving a way towards innovative design and development of nanoelectronics for advanced nanotechnology.

**Experimental Section/Methods**

***Device fabrication:***

ML-MoS$_2$ samples were synthesized on sapphire substrates by chemical vapor deposition (CVD).[22–24] Standard PMMA-assisted wet transfer technique was used to transfer MoS$_2$ single crystals from sapphire substrates to standard Si/SiO$_2$ (30 nm oxide thickness for Dev. #1-3 and 100 nm oxide thickness for Dev. #4-5) substrates using ammonia solution. PMMA residue and surface contamination was then removed by acetone/isopropanol, which was followed by N-methyl-2-pyrrolidone (NMP) solution. Electrical contacts consisting of 20 nm Bi (99.99% purity) and 50 nm Au were made by e-beam lithography and thermal evaporation.

***Device characterization:***

The electrical transport characterization of the devices was carried out using Physical Property Measurement System (PPMS) by Quantum Design in a vacuum (< 10 mTorr) cryostat, which allowed a tunable temperature ranging from 1.8 K to 400 K and a maximum tunable magnetic field of ±14 T. The device was annealed at 400 K under vacuum overnight to remove possible water and oxygen adsorbates.[18,46] All electrical transport measurements were conducted under the DC condition with the source-measuring units Keithley 2636B/2450. The gate voltage $V_{GS}$ was kept below 24 V to ensure the leakage current staying below 0.1 nA, and the $V_{GS}$ sweep rate of 0.044 V/s or slower was found to produce consistent hysteresis loops, hence the sweep rate of 0.044 V/s or slower was used for all hysteresis measurements unless otherwise specified. The optical characterization of the MoS$_2$ devices was carried out under $T$ = 4.5 K ~ 300 K temperature control with an out-of-plane magnetic field $B$ up to ±0.5 T. A 532 nm continuous-wave laser was used as an excitation source, which was focused on the sample for optical characterization with a 50× objective lens (NA = 0.5). The temperature-dependent and magnetic field-dependent Raman (1200 lines/mm) and PL (150 lines/mm) spectra were measured by a ANDOR Kymera 328i spectrometer.

The scanning tunneling microscopy (STM) studies were carried out on ML-MoS$_2$ samples, grown *in situ* by CVD on cleaved HOPG substrate, at a vacuum base pressure of $2 \times 10^{-10}$ Torr (See Figure S11). The topography measurements of the moiré superlattice patterns were obtained with a bias voltage of 0.7 V and a constant current of 2 nA.

***Extracting the hysteresis window $V_{HW}$:***
The threshold voltages $V_{th, H}$ and $V_{th, L}$ were extracted by first fitting the linear region of the HRS (forward branch) and LRS (backward branch) of the $|I_{DS}|$-*vs.*-$V_{GS}$ transfer curves respectively, then finding their corresponding x-axis intercepts. The hysteresis window $V_{HW}$ was then given by $V_{HW} \equiv V_{th, H} - V_{th, L}$ and was largely independent of $V_{DS}$, as shown in Figure S3. Another practical quantity to characterize the hysteresis loop size is the hysteresis width $(\Delta V_{th})$,[18] which is defined as the gate voltage difference between the intercepts of a small threshold current (*i.e.*, 10 nA) with the HRS and LRS $|I_{DS}|$-*vs.*-$V_{GS}$ curves so that $\Delta V_{th}$ ($|I_{DS}|$ = 10 nA) $\equiv V'_{th, H}$ ($|I_{DS}|$ = 10 nA) $- V'_{th, L}$ ($|I_{DS}|$ = 10 nA). In our study, we found that both $V_{HW}$ and $\Delta V_{th}$ ($|I_{DS}|$ = 10 nA) gave similar results while the former was independent of the choice of the threshold quantity (threshold current), as exemplified in Figure S3b. Therefore, we chose $V_{HW}$ to represent to the size of the hysteresis loop unless otherwise specified. We further remark that the Zeeman splitting under external magnetic fields would reduce the bandgap at the K or K′ valley via lowering the conduction band minimum and lifting the valence band maximum. However, this effect was on the order of $10^{-3} \sim 10^{-4}$ eV for $|B| \sim 10$ T,[47,48] which corresponded to approximately $10^{-2} \sim 10^{-3}$ V changes in the back gate voltage. Therefore, this effect was negligible in the process of extracting the threshold voltages.

## Supporting Information
Supporting Information is available from the Wiley Online Library or from the author.

## Acknowledgements
The authors are grateful to Professor Patrick A. Lee for stimulus discussions on the potential physics mechanisms for the observed hysteresis, and to Yen-Yu Lai for his valuable assistance to conducting the STM experiments. The authors thank members of TLS 09A1 at the National Synchrotron Radiation Research Center's Taiwan Light Source for their suggestions on the analysis of XPS data and prior test measurements. This work was also partially supported by the Taiwan Semiconductor Research Institute for the device fabrication, and the Instrumentation Center at National Tsing Hua University for the STEM experiments.

**Funding:** National Science Foundation (US) Institute of Quantum Information and Matter at Caltech (Award #1733907), Major Research Instrument (MRI) (Award #DMR-2117094); National Science and Technology Council (Taiwan) MOST 111-2112-M-003-008, MOST 111-2124-M-003-005, MOST 111-2628-M-003-002-MY3, NSTC 110-2634-F-009-027, NSTC 110-2112-M-A49-013-MY3, NSTC 110-2112-M-A49-022-MY2, and NSTC 112-2926-I-003-504-G; National Taiwan Normal University Yushan Fellow Distinguished Professorship; Ministry of Education (Taiwan) Yushan Fellowship

**Author contributions:** Nai-Chang Yeh, Yann-Wen Lan and Ting-Hua Lu coordinated and supervised the project. Wen-Hao Chang and Tilo H. Yang synthesized monolayer $MoS_2$ single crystals, carried out device fabrication, and performed room temperature optical characterization and basic electrical measurements. Duxing Hao carried out temperature and magnetic field dependent electrical transport measurements, where hysteretic behavior was first discovered at cryogenic temperature under finite magnetic fields. Yann-Wen Lan and Nai-Chang Yeh conceived the idea of magnetic field-induced hysteresis response; Nai-Chang Yeh proposed possible roles of low-temperature anisotropic lattice expansion and sulfur vacancies in the occurrence of hysteresis, and jointly designed further experiments with Yann-Wen Lan and Ting-Hua Lu. Wen-Hao Chang carried out synchrotron XPS experiments. Wen-Hao Chang and Yu-Chen Chang carried out low temperature Raman spectroscopic studies under the supervision of Ting-Hua Lu. Wei-Tung Liu and Naoya Kawakami carried out the STM measurements under the supervision of Chun-Liang Lin. Sheng-Zhu Ho performed the low temperature PFM measurements under the supervision of Yi-Chun Chen. Chen-Hsuan Lu carried out the PF-KPFM measurements and estimated the carrier concentrations and sulfur vacancies. Ming-Hao Liu assisted in analyzing experimental data from theoretical perspectives. Duxing Hao, Wen-Hao Chang, Yann-Wen Lan and Nai-Chang Yeh wrote the first drafts of the paper with assistance from all other authors, and Duxing Hao and Nai-Chang Yeh completed the final version of the paper.

Duxing Hao and Wen-Hao Chang contributed equally to this work.

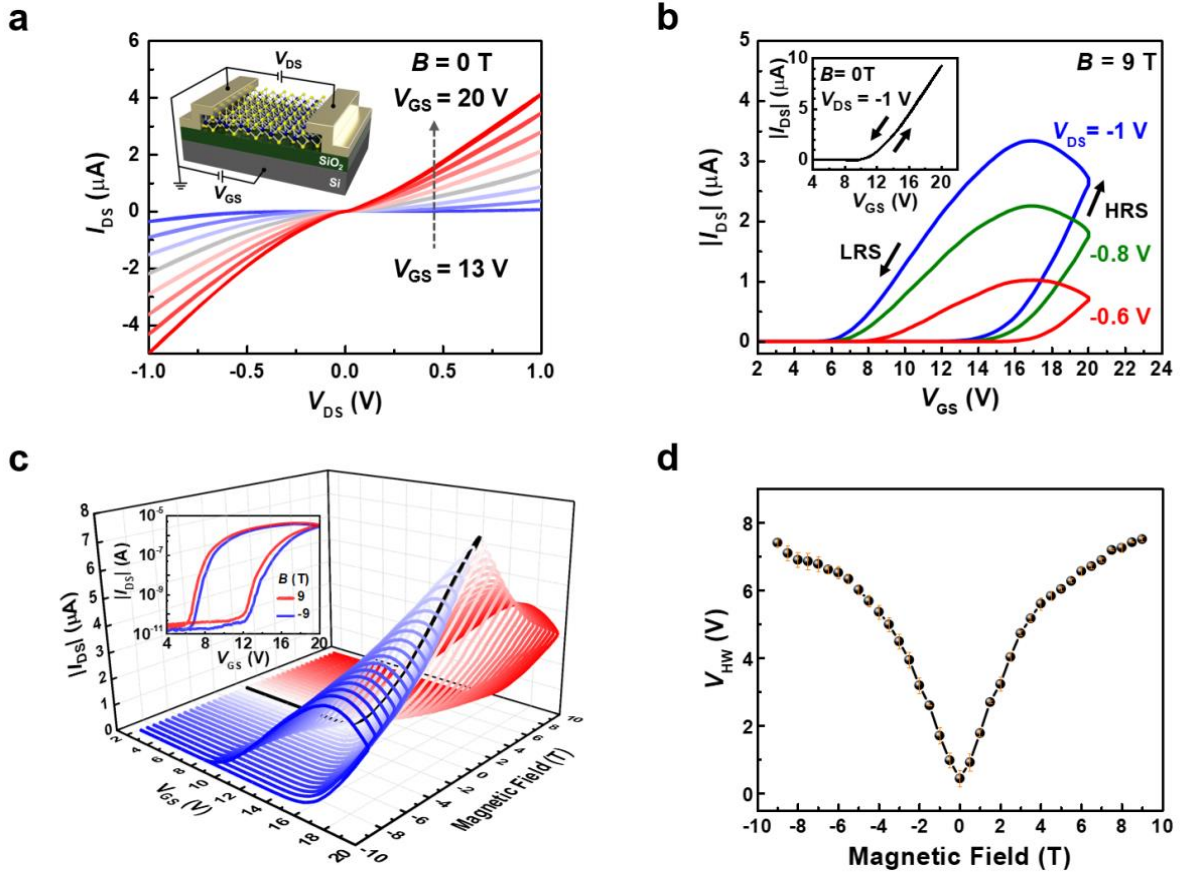

**Figure 1.** Magnetic-field induced giant hysteresis responses in ML-MoS$_2$ field-effect transistor (FET) at T = 1.8 K. a) $I_{DS}$-*vs.*-$V_{DS}$ transfer curve measured at a gate voltage ($V_{GS}$) from 13 V to 20 V with an increment of 1 V, showing largely ohmic characteristics. b) Main panel: $|I_{DS}|$-*vs.*-$V_{GS}$ hysteresis curves measured under a magnetic field of 9 T with $V_{DS}$ fixed at −1 V, −0.8 V and −0.6 V, respectively. Inset: $|I_{DS}|$-*vs.*-$V_{GS}$ curve taken at $B = 0$, showing absence of hysteresis. Black arrows in the main panel indicate the $V_{GS}$ sweeping direction for the corresponding HRS/LRS states under different $V_{DS}$ values, which reveal the counterclockwise hysteresis loops for all $V_{DS}$ values at $B = 9$ T. c) $|I_{DS}|$-*vs.*-$V_{GS}$ hysteresis under magnetic field from 9 T to −9 T with $V_{DS}$ = −1 V, showing characteristics independent of the sign of magnetic field within experimental errors, as further demonstrated in the semi-log inset for comparison of the hysteresis loops taken at $B = 9$ T (red) and − 9 T (blue). d), Hysteresis window $V_{HW}$ as a function of out-of-plane magnetic field.

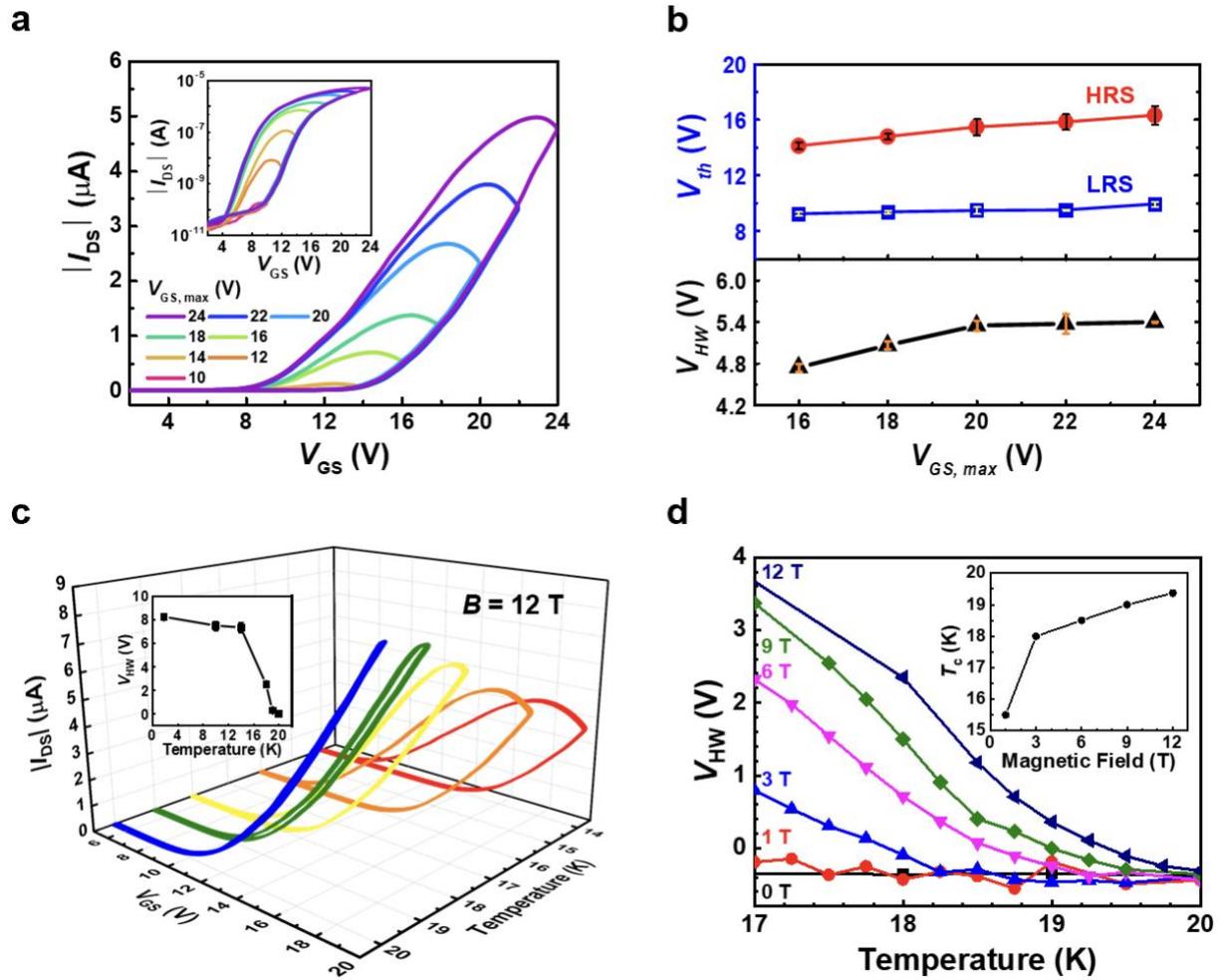

**Figure 2.** Gate voltage modulation of hysteresis and magnetic field-controlled critical temperature. a) Linear-scale (main panel) and semi-log-scale plot (inset) of $|I_{DS}|$-$vs$.-$V_{GS}$ characteristics, showing different hysteresis loop sizes under different $V_{GS}$ ranges. All measurements started at $V_{GS}$ = 2 V and ended at different $V_{GS, max}$ values under the conditions of $T$ = 10 K, $B$ = 9 T and $V_{DS}$ = −0.6 V. b) Characteristics voltages ($V_{th, H}$, $V_{th, L}$, $V_{HW}$) of the hysteresis loops in **a** are shown as a function of $V_{GS, max}$. c) Temperature dependence of $|I_{DS}|$-$vs$.-$V_{GS}$ characteristics, showing a loop closing at $T_C$ = 19.3 K under a constant magnetic field of 12 T. The inset shows the emergence of $V_{HW}$ and $P_0$ below $T_C$. d) Main panel: Determination of the critical temperature ($T_C$) from the $V_{HW}$-$T$ plots under different constant magnetic fields, where $T_C(B)$ is identified at the temperature where $V_{HW}(B, T)$ = 0. The inset shows a higher magnetic field leads to a higher $T_C$. Measurements were taken with $V_{DS}$ = −1 V.

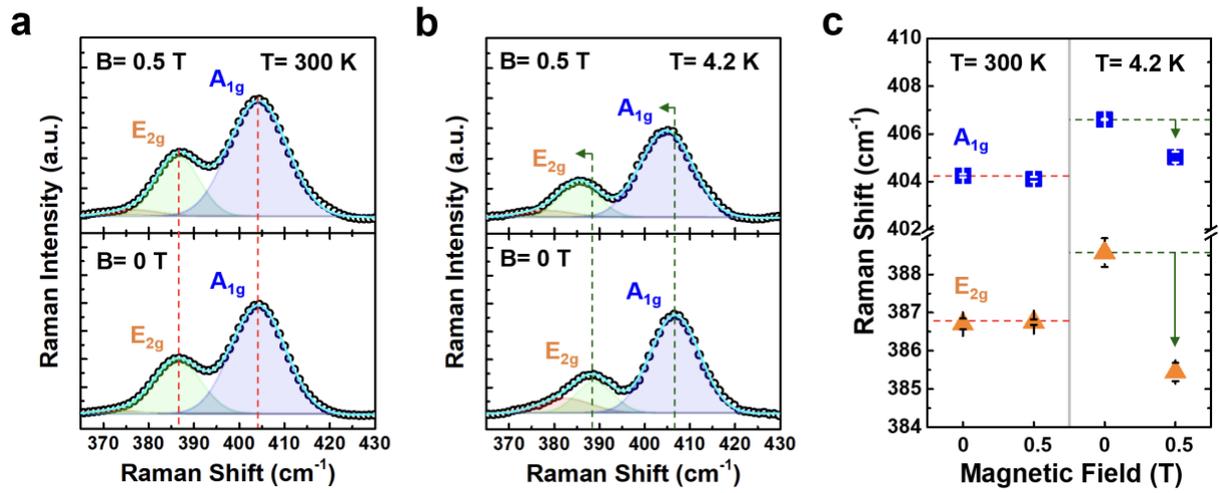

**Figure 3.** Raman characterizations of a ML-MoS$_2$ device under an out-of-plane magnetic field at 4.2 K. a, b) Raman spectra taken under $B = 0$ (bottom) and an out-of-plane magnetic field $B = 0.5$ T (top) at a) 300 K and b) 4.2 K, respectively. c) The Raman peak positions for A$_{1g}$ (blue square) and E$_{2g}$ (orange triangle) at 300 K and 4.2 K with and without magnetic field. The red (green) dashed lines indicate the peak positions of A$_{1g}$ and E$_{2g}$ at 300 K (4.2 K) under $B = 0$.

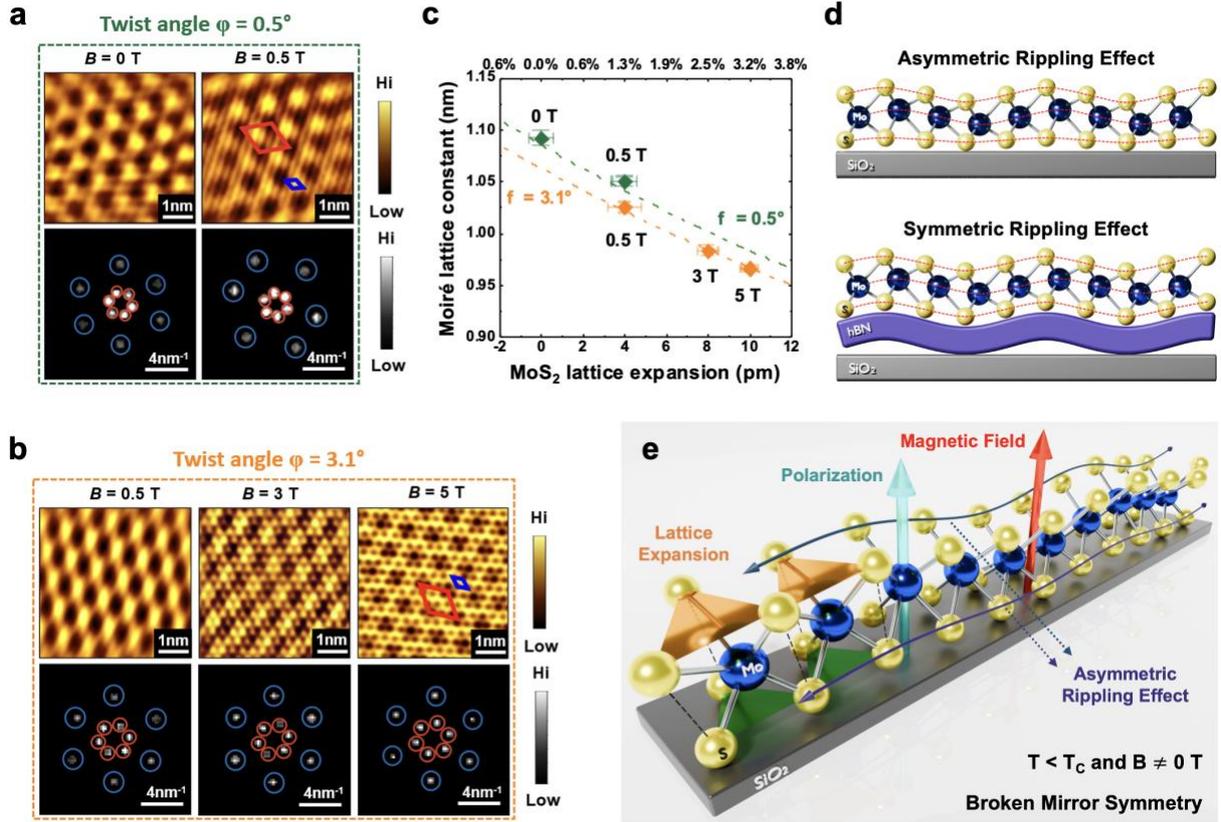

**Figure 4.** Magnetic field-induced ML-MoS$_2$ lattice expansion as characterized by scanning tunning microscopy (STM) and a proposed model for the effect of temperature and magnetic field on the structure of ML-MoS$_2$. a, b) Reconstructed moiré superlattice topography (top panels) and the corresponding filtered fast Fourier transformation (FFT; bottom panels) of ML-MoS$_2$/HOPG at 4.5 K for two (5 nm × 5 nm) regions of the same sample: a) one (5 nm × 5 nm) region with a twist angle of 0.5° between MoS$_2$ and HOPG under $B$ = 0 and 0.5 T (green); and b) the other (5 nm × 5 nm) region with a twist angle of 3.1° between MoS$_2$ and HOPG under $B$ = 0.5 T, 3.0 T and 5.0 T (orange). The red and blue rhombus in the topographic images outline the unit cell of the moiré superlattice and the MoS$_2$ lattice, respectively. Red and blue circles in the FFT images outline the corresponding reciprocal lattice sites. The STM bias voltage was fixed at 0.1 V and the tunneling current was fixed at a) 3 nA and b) 2 nA. c) The moiré superlattice periodicity of ML-MoS$_2$/HOPG and the MoS$_2$ lattice expansion (green and orange diamonds) are derived from analyzing the FFT graphs and show an excellent match to the in-plane MoS$_2$ lattice expansion theoretical model (dashed curve). The top axis shows the lattice expansion in percentage. d)**,** Schematic side-views of ML-MoS$_2$ on either a rigid SiO$_2$/Si substrate or a buffered h-BN/SiO$_2$/Si substrate and the resulting respectively asymmetric and symmetric rippling effects on the top and bottom sulfur layers. e)**,** A proposed model of asymmetric lattice expansion in ML-MoS$_2$ on SiO$_2$/Si for $T < T_C$ and under an out-of-plane

external magnetic field, leading to broken mirror symmetry that give rises to the out-of-plane polar order.

# Supporting Information
# Magnetic Field-Induced Polar Order in Monolayer
# Molybdenum Disulfide Transistors

Duxing Hao, Wen-Hao Chang, Yu-Chen Chang, Wei-Tung Liu, Sheng-Zhu Ho, Chen-Hsuan Lu, Tilo H. Yang, Naoya Kawakami, Yi-Chun Chen, Ming-Hao Liu, Chun-Liang Lin, * Ting-Hua Lu,* Yann-Wen Lan,* and Nai-Chang Yeh*

**Supplementary Figure 1. Basic characterization of devices**

Figure S1a presents an optical microscope image of one of our ML-MoS$_2$ FETs. The ML-MoS$_2$ sample in the fabricated device exhibited high degrees of homogeneity according to optical spectroscopic characterizations, including a point spectrum and a spatial map of the photoluminescence (PL) under an excitation wavelength of 532 nm as demonstrated in Figures S1b and S1c, respectively, showing a uniform PL optical band gap of $\sim 1.81$ eV. The representative single point spectrum and spatial maps of Raman spectra (Figures S1d-g) indicated a typical Raman peak separation between E$_{2g}$ ($\sim 387$ cm$^{-1}$) and A$_{1g}$ ($\sim 404$ cm$^{-1}$) of $\sim 17$ cm$^{-1}$ for ML-MoS$_2$.

The X-ray photoelectron spectroscopy (XPS) experiments, as exemplified in Figures S1h-i and Figure S2, were carried out using the scanning photoelectron microscopy (SPEM) endstation at beamline 09A1 of the National Synchrotron Radiation Research Center's Taiwan Light Source at ultra-high vacuum (UHV) conditions of $\sim 10^{-9}$ Torr near 300 K, with a pre-annealing process of sample for 20 hours at $\sim 400$ K. Monochromatic soft X-rays (400 eV photon energy) were focused down to a $\sim$200 nm spot using Fresnel-zone-plate-based focusing optics and photoelectrons are collected. These spectral and mapping characteristics of our samples after the polymer-based device fabrication process were still consistent with the values obtained from direct measurements on high quality single crystalline monolayer MoS$_2$ samples reported previously.[1,2] The spatial mapping in Figure S2 revealed no signals of magnetic impurities across the device, which corroborated with the highly uniform crystalline structure seen in scanning tunneling electron microscopy image (Figure S3) on a sample grown in the same batch.

Most FET devices studied in this work exhibited an ohmic-like low contact barrier behavior[3,4] in the $I_{DS}$-vs.-$V_{DS}$ curves from 300 K to 1.8 K under different $V_{GS}$ values, as exemplified in Figures S1g-h. The insets of Figure S4a and Figure S4b revealed an on/off ratio $>10^5$, while a shift of threshold voltage from 0 V (300 K) to 15 V (1.8 K) was apparent due to fewer thermally excited carriers at low temperature.

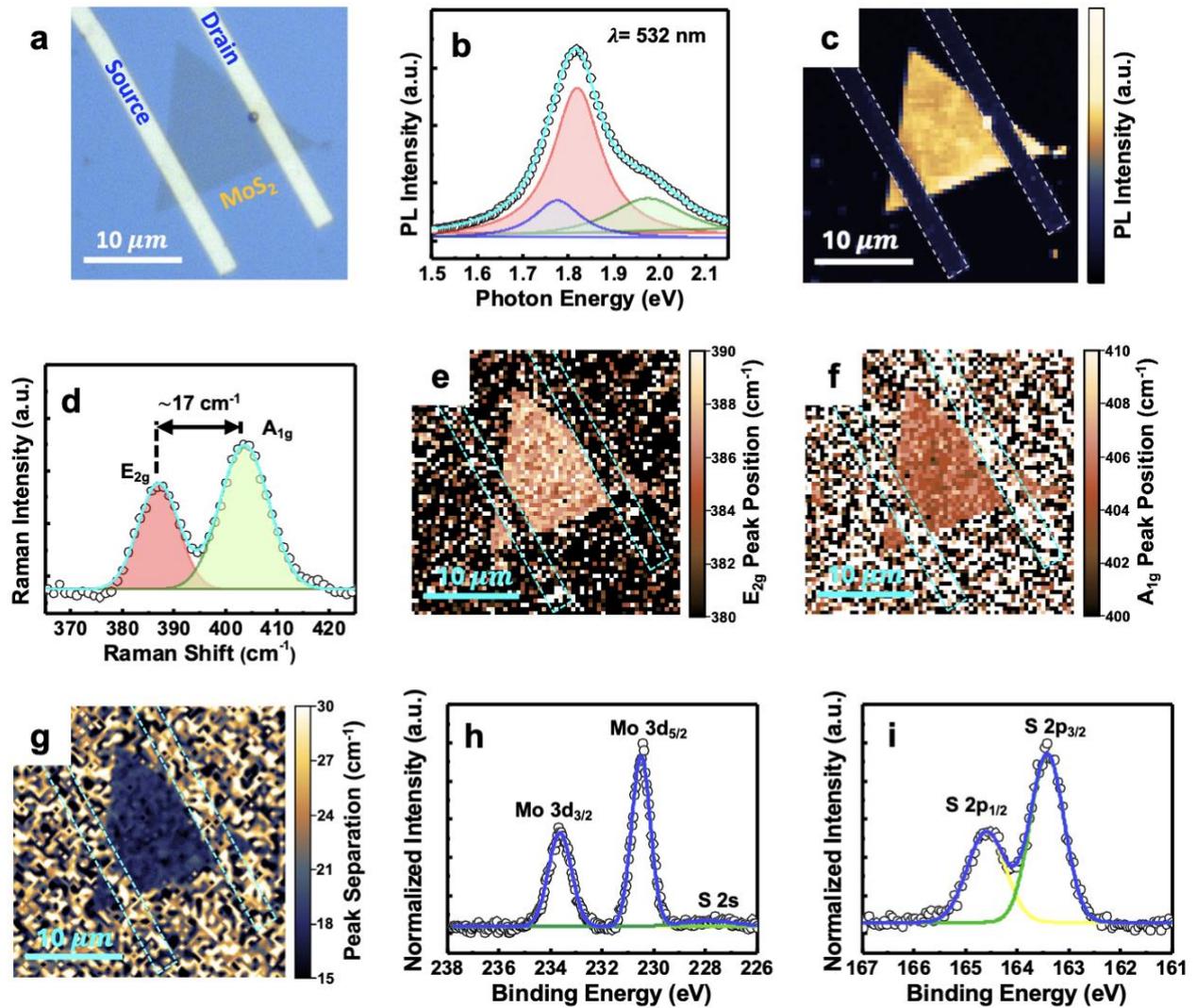

**Figure S1. Spectroscopic characterizations of the monolayer MoS₂ field-effect transistor (FET).** a) Optical microscope image of a monolayer MoS₂ device #4. b) Typical photoluminescence (PL) point spectrum and c) spatial mapping presenting an optical bandgap of ~1.81 eV. d) Typical Raman point spectrum and showing the peak position of $E_{2g}$ at 387 cm⁻¹, $A_{1g}$ at 404 cm⁻¹, and a peak separation of ~17 cm⁻¹. e) Spatial mapping of $E_{2g}$ peak position. f) Spatial mapping of $A_{1g}$ position. g) Spatial mapping of $E_{2g}$ and $A_{1g}$ peak separation. h-i) X-ray photoelectron spectroscopy (XPS) spectra revealing the binding energy of Mo 3d and S 2p, respectively. All spectroscopic measurements were performed at $T$ = 300 K.

**Supplementary Figure 2. Spatially resolved XPS measurements**

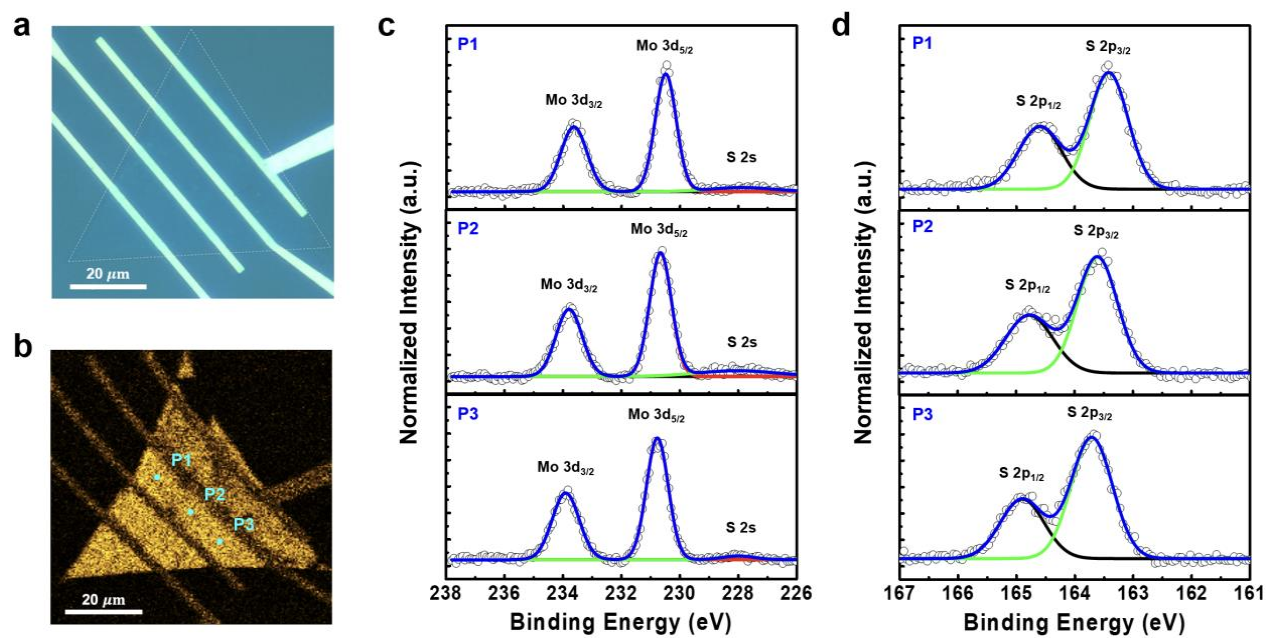

**Figure S2:** a) Optical micrograph image of the monolayer MoS₂ FET. b) The XPS Mo-3d map reveals the uniform quality of the monolayer MoS₂. c-d) The XPS spectra of the Mo 3d and S 2p, respectively, for a measured device. The locations where the point spectra were taken are indicated as light-blue dots (P1 to P3) in the mapping image of b).

**Supplementary Figure 3. Atomically resolved STEM measurements**

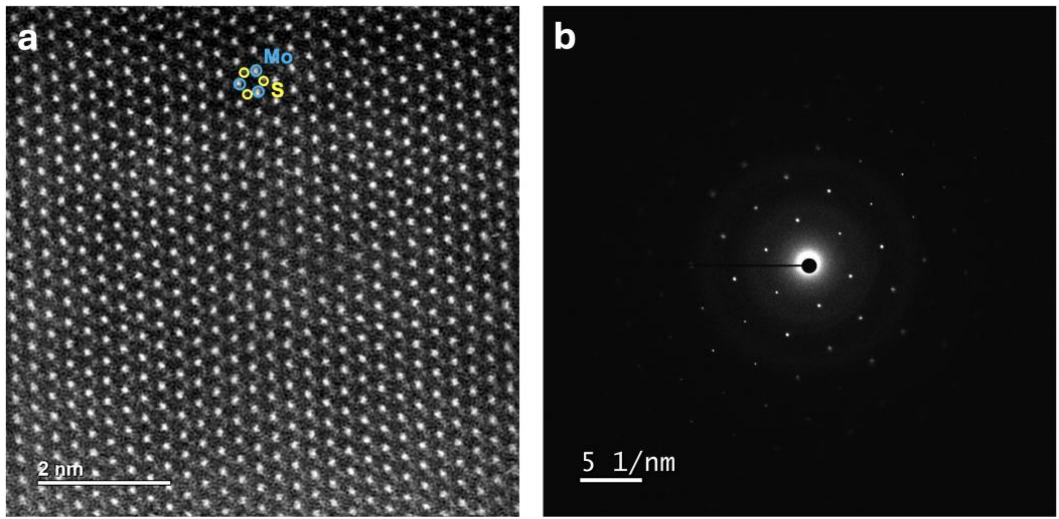

**Figure S3** a) Atomically resolved STEM image displaying an 8 nm × 8 nm area of ML MoS₂. b) Selected area electron diffraction (SAED) image further confirms the high-quality crystalline structure of the ML-MoS₂.

**Supplementary Figure 4. MoS₂ FET device characterization at 300 K and 1.8 K.**

Bi/Au contact is known for having low contact barrier and ohmic behavior down to 77 K.[4] Our study followed the same methodology and showed that our devices had an ohmic-like behavior at 300 K and a near-ohmic behavior at 1.8 K, as exemplified in Figures S4-S5.

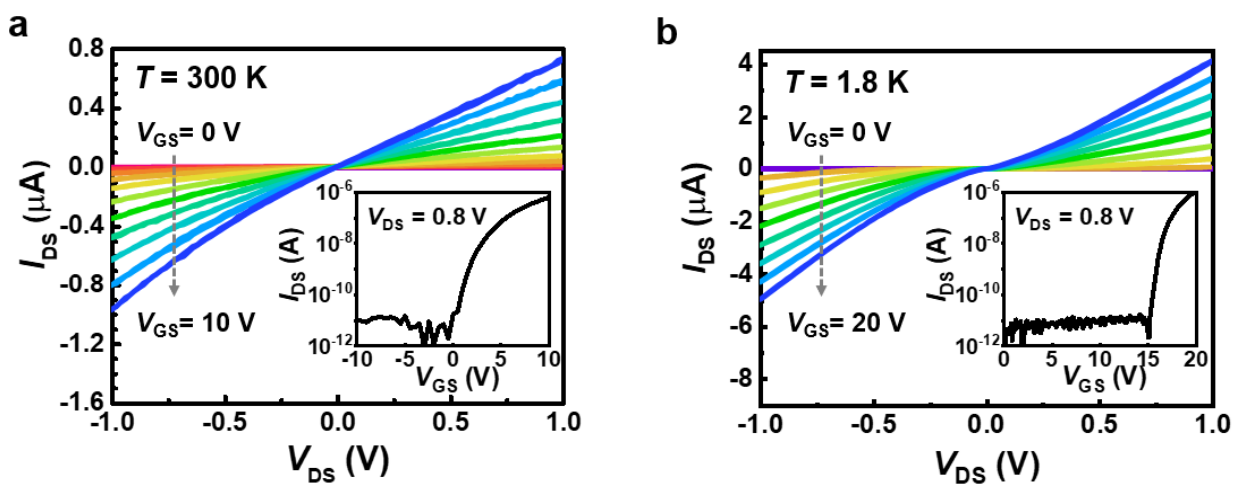

**Figure S4.** $I_{DS}$-vs.-$V_{DS}$ curves under various $V_{GS}$ measured at a) $T$ = 300 K and b) 1.8 K, respectively on device #1. Insets present logarithmic $I_{DS}$-vs.-$V_{GS}$ curves.

**Supplementary Figure 5. Ohmic-like low contact-barrier MoS₂ FET device characterization**

Under the condition that $qV_{DS} \gg k_B T$, the thermally injected current from Bi/Au contact to our MoS₂ single crystal may be approximated by the following equation:

$$I_{DS} \approx A_{2D}^* T^{1.5} \exp\left(-\frac{E_A}{k_B T}\right),$$

where $A_{2D}^* = q(8\pi k_B^3 m^*/h^2)$ is the Richardson constant for a 2D system, $E_A = q\phi_B$ is the Schottky barrier height under the flat-band condition (Figure S5a). The inset of Figure S5a showed a typical device (Device #5) that exhibited a typical Schottky barrier-like behavior, with negative linear slopes in the Arrhenius plot. In contrast, ohmic-like contacts were found in Devices #1 - #3 with extracted Schottky barrier of −20 meV, which exhibited positive slopes that asymptotically approached zero slope in the Arrhenius plot (Figure S5b) as a signature of negligible contact barrier, similar to the report by Shen *et. al.*.[4] As a result, the device showed ohmic-like $|I_{DS}|$-*vs.*-$V_{DS}$ transfer curves at 300 K (Figure S4a) and nearly-ohmic transfer curves at 1.8 K (Figure S4b). Figures S5c-d showed corresponding $I_{DS}$-*vs.*-$V_{GS}$ curve measured at 300 K and 1.8 K. The mobility for these devices was typically ~10 cm²V⁻¹S⁻¹ or better[5] and was comparable to previously reported mobility range between 0.1-10 cm²V⁻¹S⁻¹ over exfoliated ML-MoS₂ devices on typical Si/SiO₂ substrates, [6–8] which appears to be substrate Coulomb impurity-limited instead of sample quality-limited or contact-limited. [9,10]

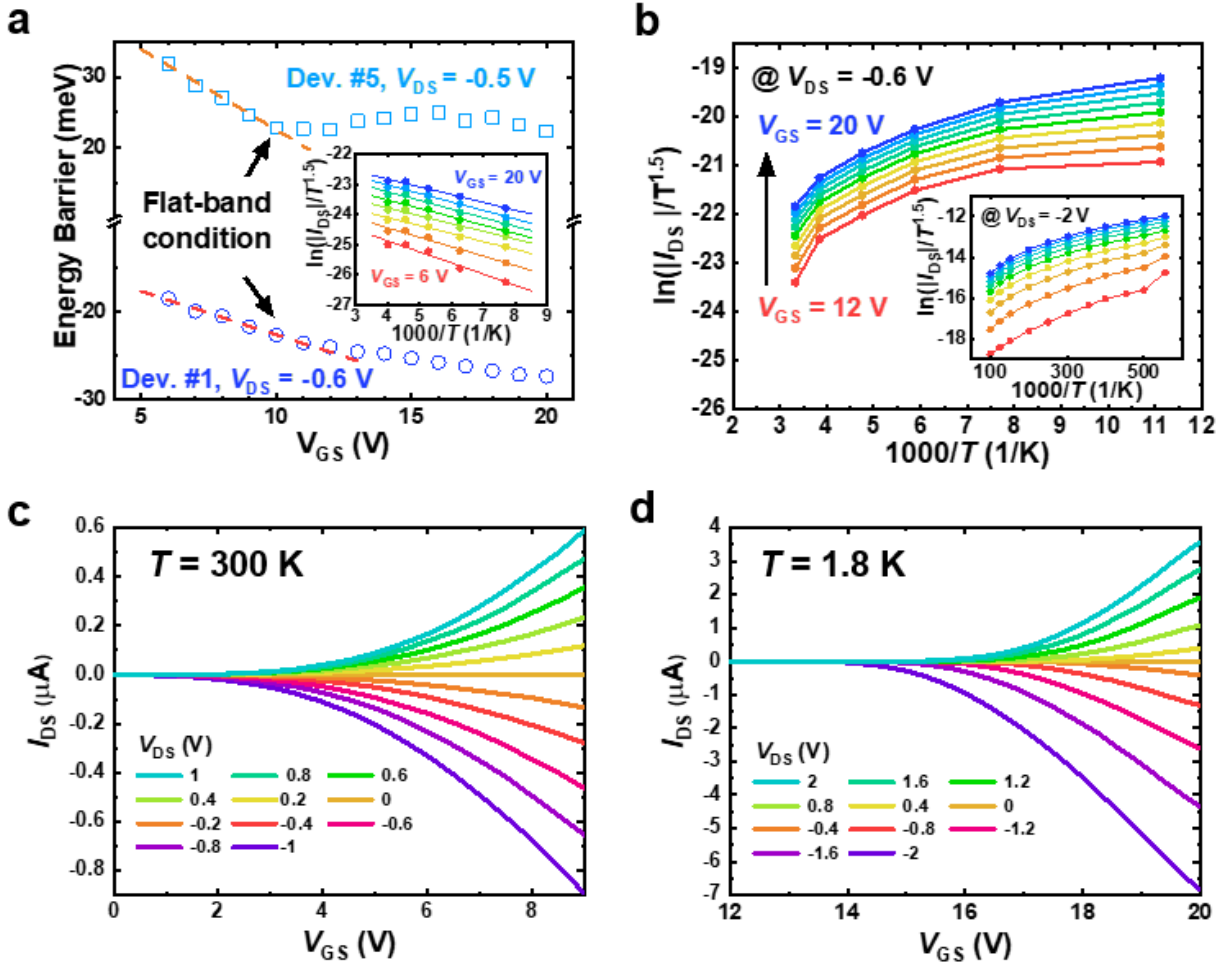

**Figure S5. Low contact barrier ML-MoS₂ FETs down to 1.8 K.** a) Main panel: Contact barrier extracted from Device #1 and Device #5 at 1.8 K. The arrows point to the dashed linear fits consistent with the flat-band conditions for ohmic contacts.[3,4] While both devices exhibited a negligible contact barrier, the high temperature dependence of Device #5 exhibited a signature linear behavior in Arrhenius plot (inset), where electron injection originated from thermionic emission across a Schottky barrier. Here $V_{GS}$ is colored from 6 V (red) to 20 V (blue) with an interval of 2V. b) Arrhenius plots of Device #1 measured between 100 K - 300 K with $V_{DS}$ = −0.6 V (main panel) and 1.8 K - 10 K (inset) with $V_{DS}$ = −2 V. Here $V_{GS}$ is colored from 12 V (red) to 20 V (blue) with an interval of 1 V. The almost saturated slope indicates a vanishing contact barrier. Similar behavior is observed on Device #2 and Device #3. c, d) $I_{DS}$-*vs.*-$V_{GS}$ characteristics of Device #1 under different $V_{DS}$ values at 300 K and 1.8 K, respectively. For all devices except Device #5, the contact resistance exhibited nearly ohmic behavior at all temperatures.

**Supplementary Note 1. Discussion on various known hysteresis-inducing mechanisms**

In this section, we aim to provide a phenomenological discussion of possible mechanisms that were previously identified in the literature and were fundamentally different from the hysteresis behavior that we observed in the $|I_{DS}|$-$vs.$-$V_{GS}$ transfer curves reported in this work.

Firstly, our findings differ drastically from previous reports of high-temperature and zero-$B$ hysteretic behavior in monolayer $MoS_2$-FETs, which have been attributed to mechanisms such as thermally-activated extrinsic or intrinsic trapped states,[11–13] absorbates,[14,15] and gate voltage-induced stress effects,[16] where clockwise hysteresis loops were observed near room temperatures without magnetic field dependences. We have also observed similar gate voltage stress type of clockwise hysteresis at higher temperature above $T_C$, as detailed in Figure S11.

At temperature below 4.2K, the presence of Schottky barrier between Cr/Au contact and $MoS_2$ can induce counterclockwise hysteresis in $|I_{DS}|$-$vs.$-$V_{GS}$ transfer curves.[17] However, this type of hysteresis bears signature of a sharp rise in $I_{DS}$ upon uncertain onset $V_{GS}$ and the $V_{HW}$ is larger with increasing $V_{GS}$ sweep rates, contrary to our observations shown in Figure 1. The Schottky barrier height in such devices is an result of a large Cr/Au contact barrier of ~ 200 meV,[17,18] whereas our devices shows negligible Schottky barrier height of ±20 meV as shown in the Figure S5a. Furthermore, our devices exhibited no abruptly changing counterclockwise hysteresis behavior at $B = 0$. These contrasts all indicate that our magnetic field-induced counterclockwise hysteresis bears different physical origin from the Schottky barrier mechanism.

Sulfur vacancies that exist in all $MoS_2$ FETs, as we discussed in the manuscript, could induce shallow donor-like trap states,[19] deep trap states above the valence band maximum,[19,20] or charge trapping at the oxide interfaces.[13,15] However, all of which would have led to magnetic field-independent clockwise hysteresis at relatively high temperatures, which contradicted to our magnetic field-dependent counterclockwise hysteresis at cryo-temperatures. In-plane motion of sulfur vacancies or defects has been shown to introduce in-plane polar order with memristor-like behavior. For example, in-plane polar order had been observed as hysteresis in the $I_{DS}$-vs.-$V_{DS}$ measurements under **in-plane** applied electric fields by introducing defects via focused ion beam,[21] chemical processing,[22] or utilizing Schottky barrier in contacts.[23] However, the scenario of ion or vacancy migration was incompatible with our observation for the following reasons: First, the densities of sulfur vacancies in our devices were very low (~ 0.2% from KPFM and STM topography measurements, see Figures. S17-S18) and no discernible magnetic impurities could be detected based on PL, Raman and XPS mapping (Figures. S1-S2). Secondly, the magnetic field-induced polar order in our devices was modulated by **out-of-plane** electric fields ($V_{GS}$), which manifested hysteresis in the $I_{DS}$-$vs.$-$V_{GS}$ measurements. In contrast, no discernable hysteresis could be found in the $I_{DS}$-vs.-$V_{DS}$ measurements with **in-plane** electric fields ($V_{DS}$), which corroborated the notion of negligible ion or vacancy migration. Finally, interlayer motion of ions or vacancies would have been energetically too costly to occur in the monolayer system at low temperatures to yield out-of-plane polarization. Thus, we may rule out vacancy-induced trap states or ion/vacancy migration as the cause of our observed magnetic-field induced hysteresis at cryo-temperatures.

Flexoelectric effect could also manifest themselves as counterclockwise hysteresis as a result of strain gradient.[24] However, such polarization relies on the strain gradient that is non-switchable, thus incompatible with our PFM results.

As for piezoelectric effect, it has been reported that the in-plane piezoelectric coefficient are much larger than the out-of-plane one,[24,25] thus less likely being the cause for our findings of $V_{HW}$ that were mostly due to out-of-plane electric field and almost independent of the in-plane electric field.

In summary, we have ruled out all known mechanisms reported in the literature and thus demonstrated the novelty of our findings of magnetic field-induced hysteresis.

**Supplementary Figure 6. Further characterization of hysteresis window**

As shown in Figure S6a, the $I_{DS}$-$vs.$-$V_{GS}$ transfer curves were taken under different source-drain bias voltages $V_{DS}$ at $T = 1.8$ K, $B = -9$ T. The hysteresis window was extracted using both the threshold voltage ($V_{HW}$) and threshold current ($\Delta V_{th}$, $|I_{DS}| = 10$ nA) respectively, as shown in Figure S6b. Two extraction methods showed similar results and they both showed no significant $V_{DS}$ bias voltage dependence.

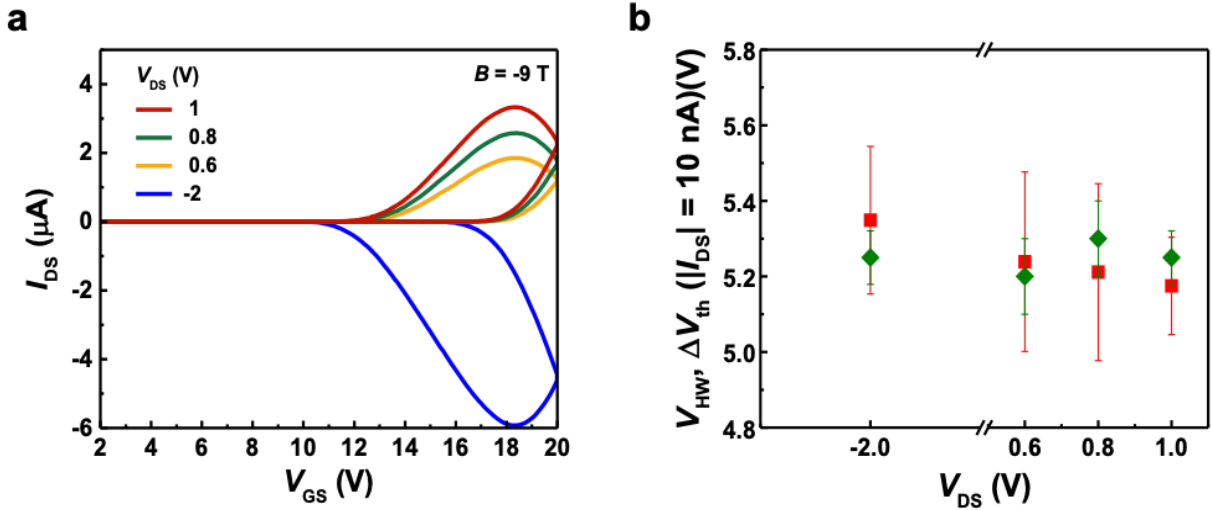

**Figure S6. $V_{DS}$-independent hysteresis loop size.** a) $I_{DS}$-$vs.$-$V_{GS}$ hysteresis curves measured under a magnetic field of $-9$ T with $V_{DS}$ fixed at 1 V, 0.8 V, 0.6 V and $-2$V, respectively. b) Extracted $V_{HW}$ (green) and $\Delta V_{th}$ ($|I_{DS}| = 10$ nA) (red).

**Supplementary Figure 7.** Frequency response of the hysteresis window

The frequency response of the hysteresis window is also studied and shown in Figure S7. At the magnetic field of 9 T, four cycles of the $|I_{DS}|$-$vs.$-$V_{GS}$ hysteresis curve were taken with different $V_{GS}$ sweep rates. The extracted $V_{HW}$ as a function of the cycle number is plotted. In the fast sweep limit ($e.g.$, 2.178 V/s), it would take one or two cycles to ramp up to a stable $V_{HW}$, and this stable $V_{HW}$ was smaller than that achieved with a much slower $V_{GS}$ sweep rate. This behavior may be due to the slower response time for the electric polarization at low temperatures.

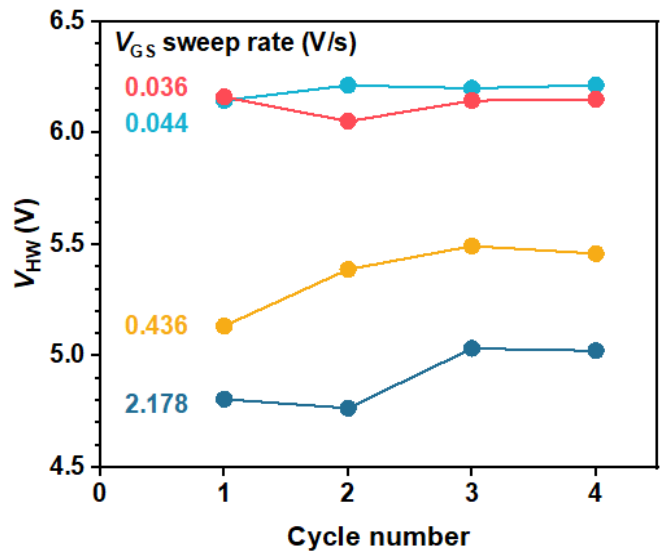

**Figure S7. Frequency response of the hysteresis window.** The extracted $V_{HW}$ measured at a different $V_{GS}$ sweep rate, each measured four times in a consecutive manner. $V_{GS}$ range is 2-24 Volts while the sweep rate varies by a combination of voltage step and interval time between voltage steps. The measurement was taken at $B$ = 9 T, $T$ = 1.8 K and $V_{DS}$ = 2V on Device #1.

**Supplementary Figure 8. Low-temperature magnetic field-dependent electronic transport data on additional devices**

Figure S8 shows consistent high and low field behavior like Figure 1 in the main text on two additional devices. Figure S8a shows the $|I_{DS}|$-$vs.$-$V_{GS}$ hysteresis under magnetic field from 9 T to −9 T with $V_{DS}$ = 0.4 V. The inset shows $|I_{DS}|$-$vs.$-$V_{GS}$ hysteresis in the semi-log scale and it is clear this hysteresis is symmetric with respect to direction change of the out-of-plane magnetic field. Figures S8b-c shows the $V_{HW}$, $V_{th, H}$ and $V_{th, L}$ extracted from the main panel of Figure S8a. Figures S8d-f showing this hysteresis are sensitive to low fields as small as 100 mT.

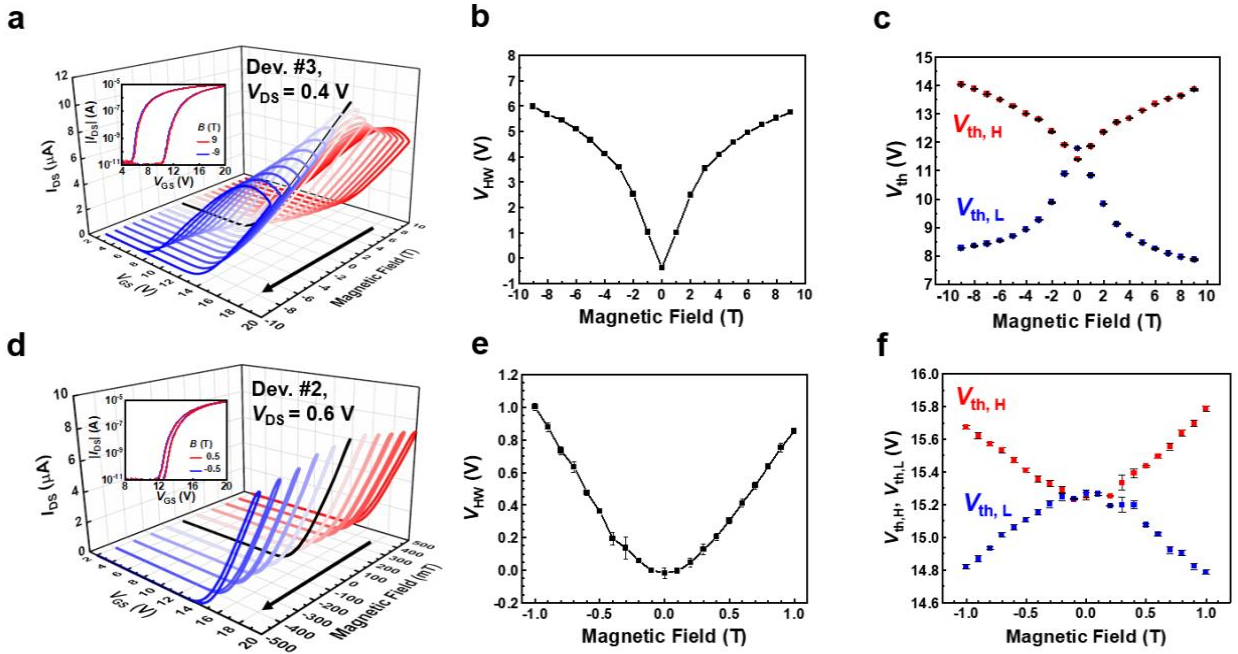

**Figure S8. Consistent high and low-field hysteresis behavior measured on additional devices.** a) Main panel: $|I_{DS}|$-$vs.$-$V_{GS}$ hysteresis under magnetic field from 9 T to -9 T (sweep direction indicated by the black arrow) with $V_{DS}$ = 0.4 V measured on Device #3. A magnetic field sign-symmetry under $B$ = ±9 T is presented within experimental error as shown in the semi-log inset. b) Magnetic field-controlled $V_{HW}$ extracted from a). c) Threshold gate voltages $V_{th, H}$ and $V_{th, L}$ for Device #3 are shown as a function of the applied magnetic field from −9 T to 9 T. d) Main panel: $|I_{DS}|$-$vs.$-$V_{GS}$ hysteresis loops under low magnetic fields from $B$ = −0.5 T to 0.5 T (sweep direction indicated by the black arrow) with $V_{DS}$ = 0.6 $V$ measured on Device #2 are shown for the sake of clarity. A magnetic field sign-symmetric low-field hysteresis behavior under $B$ = ±0.5 T is presented in the semi-log inset. e) Magnetic field controlled $V_{HW}$ extracted from d). f) Threshold gate voltages $V_{th, H}$ and $V_{th, L}$ for Device #2 are shown as a function of the applied magnetic field from −1.0 T to 1.0 T.

**Supplementary Figure 9. Magnetoresistance measurements**

Magnetoresistance measurements of Device #5 are shown in Figure S9. The measurement started from $B = 0$, ramping up the gate voltage to 20 V and holding for the gate voltage for 10 minutes, which was followed by a magnetic field sweep sequence of $B = 0$ T→ 14 T → 0 T→ −14 T → 0 T with a ramping rate of 10 mT/sec while the resistance was taken. No significant hysteresis was seen in consecutive four cycles; hence the data shown were averaged.

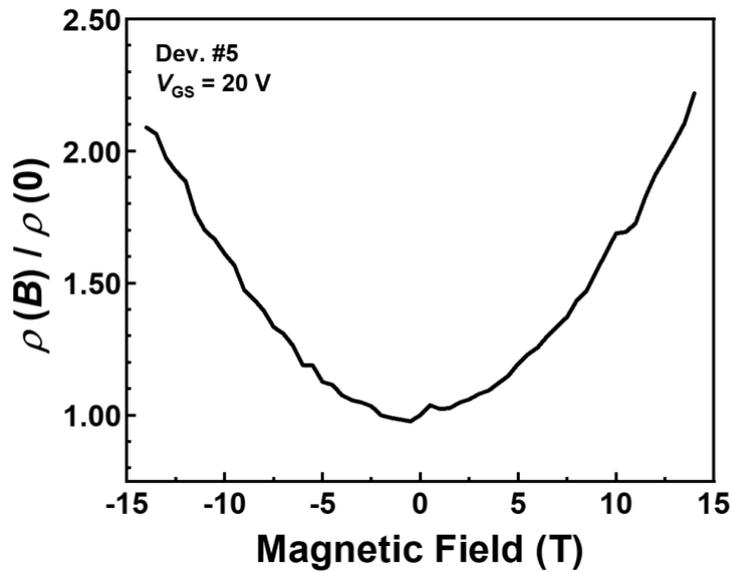

**Figure S9. Magnetoresistance measured on Dev. #5 at 1.8 K.** The was obtained by sweeping the gate voltage up to 20 V at B =0 T with $V_{DS}$ = -2 V and hold for 10 minutes. Then magnetic field was ramped (10 mT/s) to the positive maximum of 14 T followed by negative maximum of -14 T then zero, finishing one cycle. Four consecutive magnetic field ramping cycle was tested and no significant hysteresis was seen. The data was then averaged and shown.

**Supplementary Figure 10. Magnetic field sign-symmetric hysteretic behavior**

Figure S10 shows a reversed magnetic field sweep right after measurement of Figure 1c, where the only thing changed was the magnetic field ramping direction. It is shown that within the experimental error, $V_{HW}$ does not have a magnetic hysteresis as seen in Figure S10b.

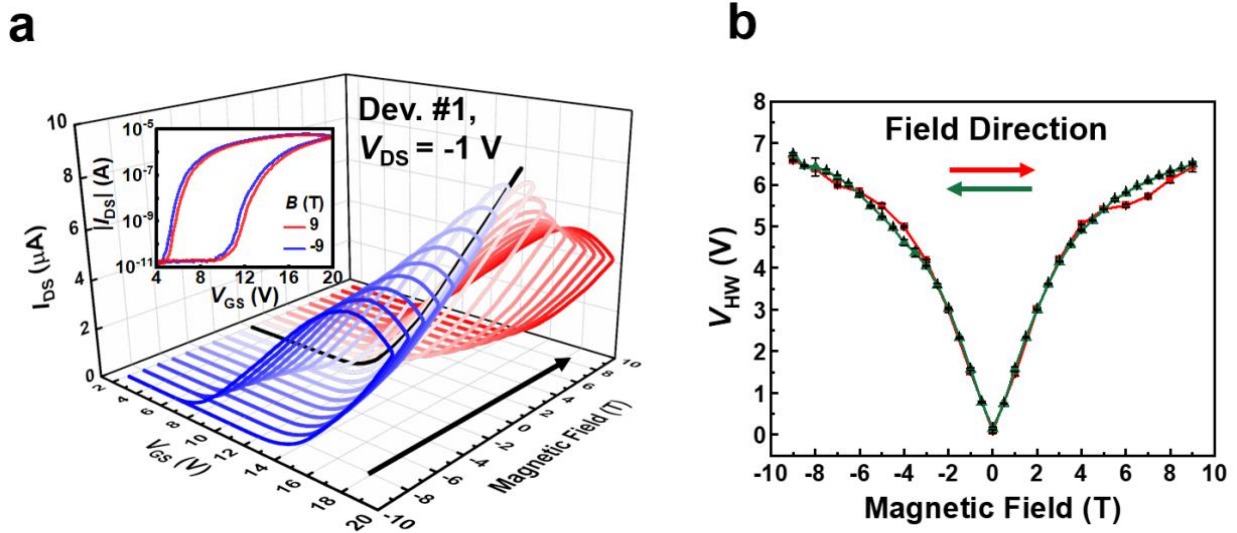

**Figure S10. Magnetic field sign-symmetric hysteretic behavior of the monolayer MoS₂ FET devices.** a) $|I_{DS}|$-*vs.*-$V_{GS}$ hysteresis under magnetic field from −9 T to 9 T (sweep direction indicated by the black arrow) with $V_{DS} = −1$ V measured after Figure 1c. A magnetic field sign-symmetry under $B = ±9$ T is presented within experimental error as shown in the semi-log inset. b) Magnetic-field controlled $V_{HW}$ and extracted from a) and Figure 1c. Red and green arrows indicate the corresponding field ramping directions. These results strongly suggest the absence of any discernible out-of-plane magnetization.

**Supplementary Figure 11. Magnetic field-independent clockwise hysteresis above $T_C$**

Above $T_C$, thermally activated and magnetic field-independent small **negative** $V_{HW}$ (clockwise hysteresis) was observed due to gate voltage[16] stress and $SiO_2$-$MoS_2$ interfacial trapping states,[11,12] as exemplified in Figure S11a. In contrast, below $T_C$, magnetic field-induced large **positive** $V_{HW}$ (counterclockwise hysteresis) was observed to be increasing at lower temperature, as shown in Figure S11b.

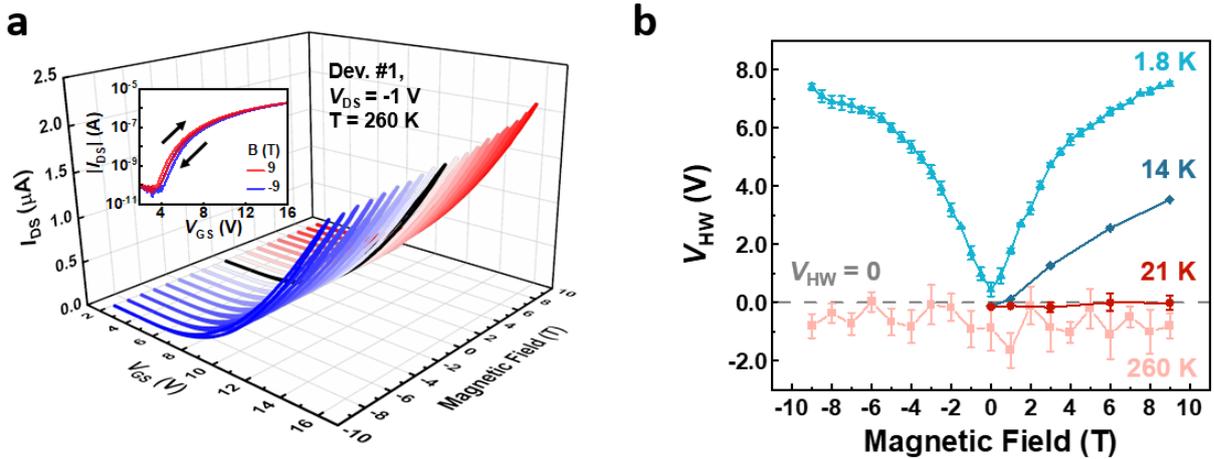

**Figure S11. Magnetic field-independent clockwise hysteresis at 260 K.** a) $|I_{DS}|$-$vs.$-$V_{GS}$ hysteresis under magnetic field from 9 T to −9 T with $V_{DS}$ = −1 V measured at 260 K. As exemplified in the semi-log inset, clockwise hysteresis loop (indicated by black arrow) that was almost independent of magnetic field was observed under $B$ = 0 and $B$ = −9 T. b) $V_{HW}$-$vs.$-$B$ taken at 1.8 K, 14 K, 21 K and 260 K. Grey dashed line shows the constant zero $V_{HW}$.

**Supplementary Figure 12. Additional temperature-dependent measurements**

Figure S12 shows measurements like Figures 2c-d on additional devices. Despite the fact that Device #5 was fabricated on a different batch and suffered from a positive Contact barrier on its contacts, we found similar (although nosier) temperature-dependent behavior to those observed in ohmic-like Devices #1-#3. As summarized in Figure S12c, comparable $T_C$ ($B$) values within ±1 K were found.

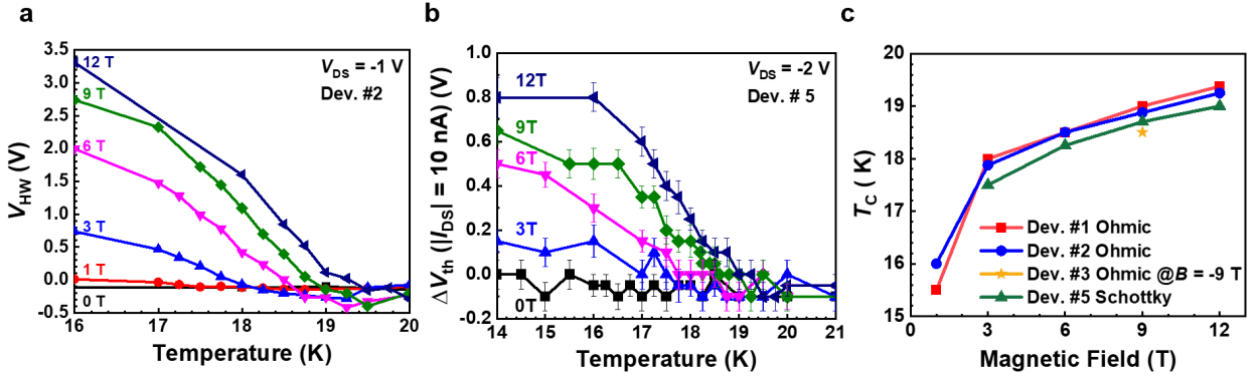

**Figure S12. Consistent temperature-dependent behavior on additional devices.** a) Temperature dependence of $V_{HW}$ under different magnetic fields from Device #2. b) Temperature dependence of $\Delta V_{th}$ ($|I_{DS}|$ = 10 nA) under different magnetic fields from Device #5. c) Magnetic field-dependent $T_C$ values for Devices #1, #2, #3 and #5. All $T_C$ values were measured under positive magnetic fields except Device #3 where the $T_C$ value was measured at $B = -9$ T.

**Supplementary Figure 13.  Temperature dependent Raman spectroscopy**

Temperature dependent Raman spectroscopy result taken under $B = 0$ is shown in Figure S10a, where the $E_{2g}$ and $A_{1g}$ Raman modes of $MoS_2$ and a Raman mode of Si were extracted as shown in Figure S13b. The green dashed line shows the linear fitting at temperatures higher than 100K, yielding a first-order temperature coefficient of $-0.0106$ cm$^{-1}$/K and $-0.0114$ cm$^{-1}$/K for $A_{1g}$ and $E_{2g}$ mode, respectively. These values are similar to previous studies.[26–28]

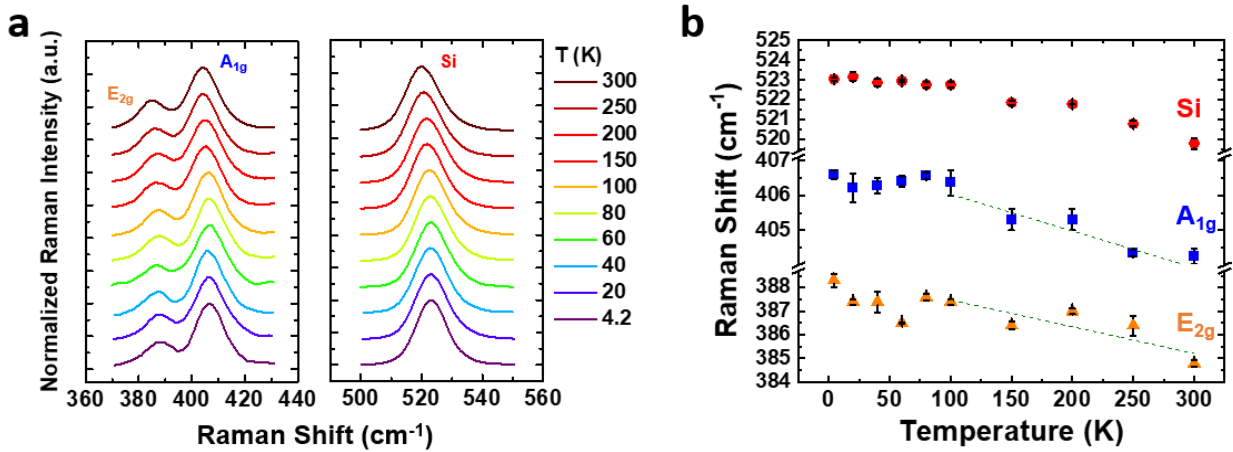

**Figure S13. Temperature-dependent Raman spectra and extracted shifts of the Raman modes under $B = 0$**. a) Temperature dependence of the monolayer $MoS_2$ and Si-substrate Raman spectra from $T = 300$ K to 4.2 K. b) Extracted temperature dependence of the Si (dot), $A_{1g}$ (square) and $E_{2g}$ (triangle) Raman peak positions. The green dashed lines are fitted based on 100 K to 300 K data.

**Supplementary Figure 14. Scanning tunneling microscopy on HOPG/MoS₂ moiré superlattice**

The ML-MoS₂ sample for STM measurements was synthesized by CVD on an *in situ* cleaved HOPG substrate. The as-grown sample was then outgassed *in situ* at a temperature of 800 K and a vacuum of $2 \times 10^{-10}$ Torr for an hour before measurements. The ML-MoS₂/HOPG sample was then transferred to an STM chamber and measured at 4.5 K under various magnetic fields using an electrochemical-etched tungsten tip, whose quality was verified by test measurements on Au (111) surface states. Further STM calibration was done by scanning on pure HOPG areas of the sample at 4.5 K prior to the study of sample area covered by a ML-MoS₂. Due to the lattice constant mismatch as well as a small twist angle between HOPG and the as-grown ML-MoS₂, moiré superlattice patterns were observed in the STM topography. As shown in Figures S11a-f, a bias voltage-dependent moiré topography was observed due to modified electronic coupling between HOPG and MoS₂, similar to the findings in ref.[29] However, this dependence only changed the relative intensity rather than the position of the lattice points, as long as the Fermi level was kept within the band gap of MoS₂. Eventually, at a sufficiently large bias voltage of 0.7 V, a clear topography of the top-layer sulfur atoms emerged without moiré patterns (Figures S14g-h). Therefore, a bias voltage of 0.7 V and a constant current of 2 nA were used to obtain the topography unless otherwise specified.

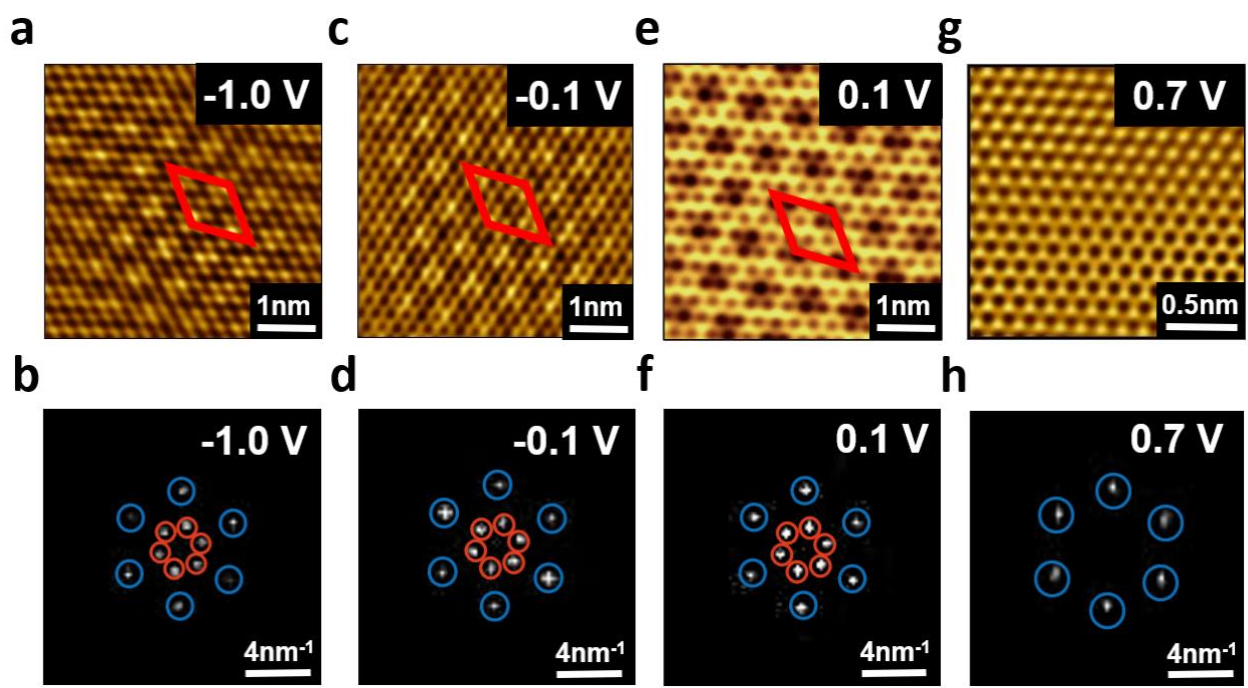

**Figure S14. Bias voltage-dependent moiré patterns.** a-f) Reconstructed moiré superlattice topography and their filtered FFT images measured under a bias voltage of a, b) −1 V, c, d) −0.1 V and e, f) 0.1 V, respectively. g, h) Measured topography of $MoS_2$ at a bias voltage of 0.7 V, showing mainly top sulfur atoms of $MoS_2$.

**Supplementary Figure 15. Extracting the ML-MoS₂ lattice expansion from moiré patterns**

Moiré patterns are very sensitive to the lattice mismatch between ML-MoS₂ and the underlying HOPG, hence served as an excellent tool to accurately determined the MoS₂ lattice expansion by studying the STM topography of the same sample area under various constant magnetic fields. As exemplified in Figures S15a-d, fast Fourier transformation (FFT) was performed over the obtained raw topographic image, and then the MoS₂ lattice constant, moiré periodicity and twist angle were derived from studying the FFT pattern. Inverse FFT was performed on the filtered FFT image to obtain the filtered topography that highlighted the moiré pattern evolution under magnetic field. These results unambiguously demonstrated that out-of-plane magnetic fields induced lattice expansion in ML-MoS₂ because the lattice constant of HOPG was known to have no magnetic field dependence from previous study.[30] Thus, under the FET configuration, ML-MoS₂ was expected to undergo similar magnetic field-induced lattice expansion, which would give rise to strain and thus the asymmetric rippling effects between the top and bottom sulfur layers. This asymmetric lattice expansion ultimately led to the emergence of ferroelectric-like polar order below the critical temperature and appeared as tunable counterclockwise hysteresis in FET measurements.

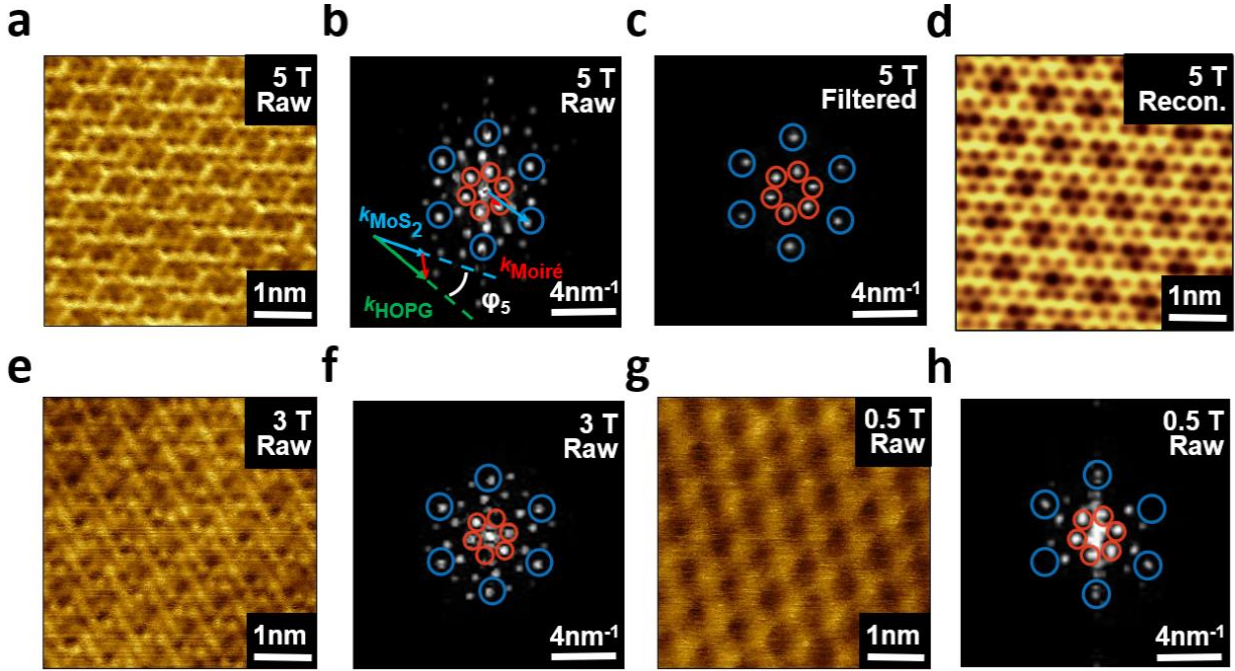

**Figure S15. Image processing and raw data of STM topography.** a) Raw data of ML-MoS₂/HOPG moiré superlattice topography under a magnetic field of 5T. b) FFT image of the topography shown in **a**), where blue and red circles highlighted the MoS₂ reciprocal lattice

vectors and the moiré reciprocal superlattice vectors, respectively. Red, blue, and green arrows on the side are exaggerated illustrations of the reciprocal lattice vectors of the moiré superlattice, MoS$_2$ lattice and HOPG lattice, respectively, where the twisted angle was measured between the MoS$_2$ and HOPG reciprocal lattice vectors. c) Filtered FFT image. d) Topography reconstructed from the inverse FFT on c). e, f) Raw data of moiré superlattice topography and FFT image measured under $B = 3$ T, respectively. g, h), Raw data of moiré superlattice topography and FFT image measured under $B = 0.5$ T, respectively. Here we note that the other signal points that were present in the raw data of FFT images of b), f) and h) but were filtered out in our analyses originated from the increased scanning signal resolution in STM after applying a strong magnetic field, which not only enhanced the brightness of each lattice point in reciprocal space, but also made new convolution signals extending outward from lattice points more apparent. Therefore, the extra points extending outward from the MoS$_2$ periodic reciprocal space vector correspond to the MoS$_2$ convolution signals, and the extra points extending from the blue outer ring of the moiré periodic reciprocal space vector corresponding to the moiré convolution signals.

**Theoretical modeling for deriving the MoS$_2$ lattice expansion from varying moiré patterns**

Since the lattice constant of HOPG remains constant under magnetic field,[30] the lattice match ($\delta$) caused by magnetic field may be solely attributed to the changes in the ML-MoS$_2$ lattice constant so that $\delta = (a_G + a_M + \Delta a) / a_G$, where $a_G = 0.246$ nm and $a_M = 0.318$ nm are the lattice constant of HOPG and ML-MoS$_2$ under zero magnetic field, respectively, and $\Delta a$ is the ML-MoS$_2$ lattice expansion under a finite magnetic field.[31] Given the twist angle ($\varphi$) between HOPG and MoS$_2$, the expected moiré pattern periodicity ($\lambda$) becomes $\lambda = \dfrac{(1+\delta)a_G}{\sqrt{2(1+\delta)(1-\cos\varphi) + \delta^2}}$.

**Supplementary Figure 16. Low temperature piezo-response force microscopy (PFM) measurements**

To study the hysteresis behavior of ML-MoS$_2$ flakes on SiO$_2$/Si substrate under magnetic fields, piezo-response force microscopy (PFM) measurements were carried out at magnetic field strengths of $B = 0$ and $B = 3$ T at a temperature of 1.6 K, as shown in Figure S16. The ferroelectric properties were analyzed using off-field piezo-response hysteresis loops to remove the electrostatic contributions. The PFM signals measured at B=3 T displays the characteristic ferroelectric polarization switching through the butterfly hysteresis loops of PFM amplitude (blue) and square phase (red) loops against the bias voltage. These results provide evidence for the presence of magnetic field-induced out-of-plane ferroelectricity in the ML-MoS$_2$ and demonstrate local point switching upon contact with a biased tip, as evidenced by the PFM hysteresis loops. Note that unlike the single polarity of the flexoelectric dipole induced by strain gradients, the ferroelectric hysteresis exhibits an opposite polarization state under a reversed field.

The hysteresis measurements using piezo-response force microscopy (PFM) were conducted utilizing a commercial cryogenic scanning probe microscope system (attoAFM I, Attocube) equipped with a closed-cycle cryostat (attoDRY 2100 with 9 T magnet, Attocube) operating at 1.6 K. A commercial platinum silicide (PtSi) coated tips with a spring constant of 2.8 N/m (NANOSENSORS PtSi-FM) was used to assess hysteresis, driven by a $V_{RMS} = 1.5$ V ac voltage at a contact-resonance frequency of about 300 kHz. Off-field hysteresis loops were obtained by switching spectroscopic techniques under pulse sequences generated by an arbitrary waveform generator (G5100A, Picotest).

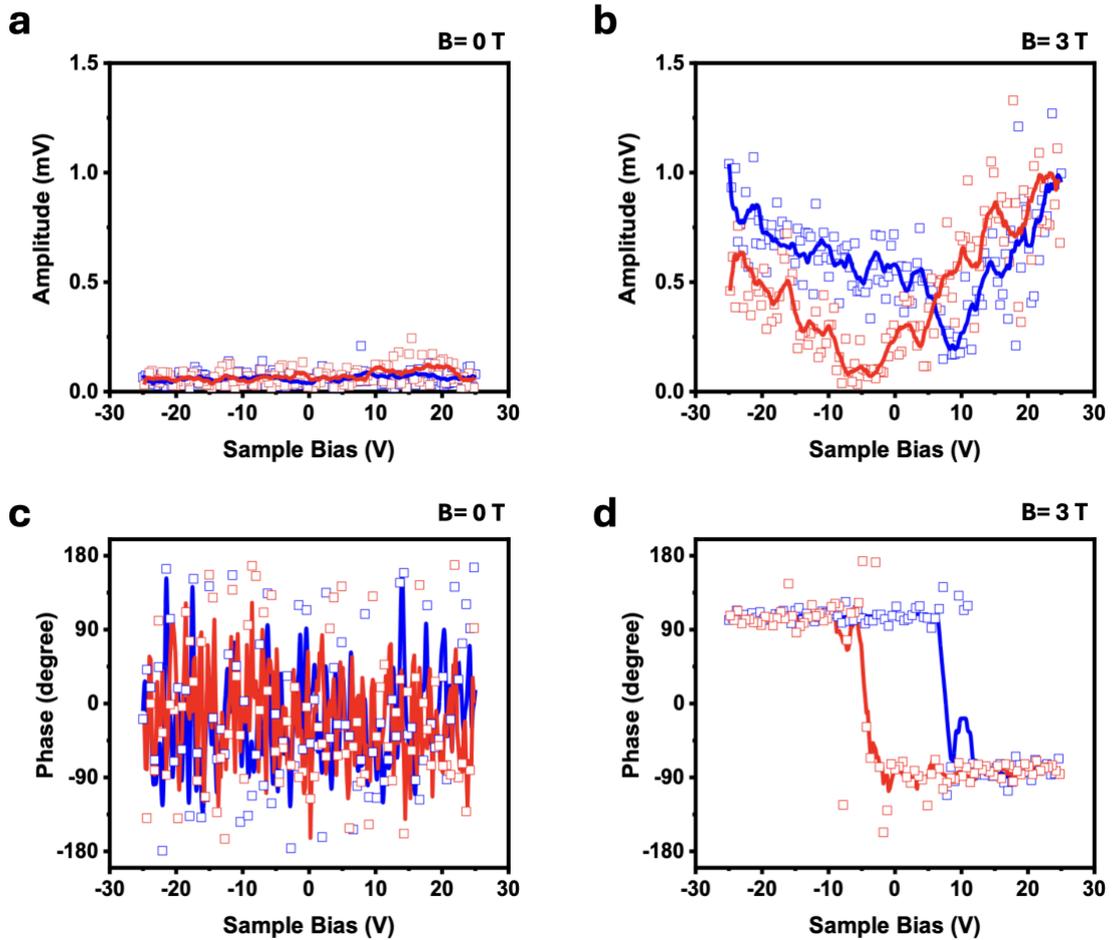

**Figure S16. Low temperature piezo-response force microscopy (PFM) measurements of ML-MoS₂ at different magnetic fields**. a, b) The amplitude and c, d) the phase of the PFM, measured at $B = 0$ and $B = 3$ T, respectively. Blue and red curves represent the response when voltage is applied from negative to positive and when reversed from positive to negative, respectively.

**Supplementary Figure 17. Characterizing the sulfur vacancy concentration**

The sulfur vacancies in ML-MoS$_2$ may play an important role in our devices. Therefore, a typical FET device (Dev. #5), which we believed to possess representative sulfur vacancy concentrations, was studied using Kelvin probe force microscopy (KPFM), as shown in Figure S14. We also presented in Figure S17 direct surface topography imaging of one ML-MoS$_2$ sample by STM, which revealed a sulfur vacancy level similar to the low density of sulfur vacancies ~0.2% (Dev. #1-5) found in the ML-MoS$_2$ FET devices. The details are as follows.

The work function of the ML-MoS$_2$ was measured by the Peak Force Kelvin Probe Force Microscopy (PF-KPFM) calibrated with respect to the work function of gold at 4.82 eV. The contact potential difference (CPD) between the tip and the sample is given by $\Delta V_{\text{CPD}} = \phi_{\text{sample}} - \phi_{\text{tip}}$, where $\phi$ is the work function. Therefore, the work function of the ML-MoS$_2$ sample becomes:

$$\phi_{\text{MoS}_2} = \phi_{\text{tip}} + \Delta V_{\text{CPD}}^{\text{MoS}_2} = \phi_{\text{Au}} + \Delta V_{\text{CPD}}^{\text{MoS}_2} - \Delta V_{\text{CPD}}^{\text{Au}}$$

where $\Delta V_{\text{CPD}}^{\text{MoS}_2}$ was measured to be $0.52 \pm 0.23$ V, while $\Delta V_{\text{CPD}}^{\text{Au}}$ was measured to be $0.39 \pm 0.22$ V in a scan size of (1 μm × 1 μm, 512*512 grid) area on the fabricated Device #5. The Fermi level $E_{\text{F}}$ is therefore determined from $\phi_{\text{MoS}_2} \sim$ 4.9eV, which is located 0.4 eV above at the intrinsic Fermi level ($E_{\text{i}}$) of ML-MoS$_2$, as shown in the Figure S14. Here we note that approximation of $E_{\text{i}}$ was used as $E_{\text{i}}$ can be expressed as $E_{\text{i}} = (E_{\text{c}} + E_{\text{v}} + k_{\text{B}}T \ln(m_{\text{p}}^*/m_{\text{n}}^*))/2 \approx (E_c + E_v)/2$, where $m_{\text{p}}^*$ and $m_{\text{n}}^*$ are the effective mass of holes and electrons of MoS$_2$, respectively, and the $k_{\text{B}}T \ln(m_{\text{p}}^*/m_{\text{n}}^*)$ term is negligible comparing to ($E_c + E_v$). The electron concentration of MoS$_2$ could be calculated with the following formula:

$$n = n_{\text{i}} \exp(\frac{E_{\text{F}} - E_{\text{i}}}{k_{\text{B}}T})$$

where $k_{\text{B}}$ is Boltzmann constant, $T$ is the temperature, $n_{\text{i}}$, are the intrinsic electron concentration of MoS$_2$. With $n_{\text{i}} \sim 10^6$ cm$^{-2}$ at room temperature[32] and measured $E_{\text{F}} - E_{\text{i}} = 0.4$ eV, the corresponding electron concentration of the MoS$_2$ sample was estimated as $n \approx 4.8 \times 10^{12}$ cm$^{-2}$. Assuming those electrons were induced by sulfur vacancies, the order of magnitude of sulfur vacancy population density can be estimated as $na^2\sqrt{3}/4 \sim 0.2\%$, where $a = 0.318$ nm is the lattice constant of MoS$_2$.

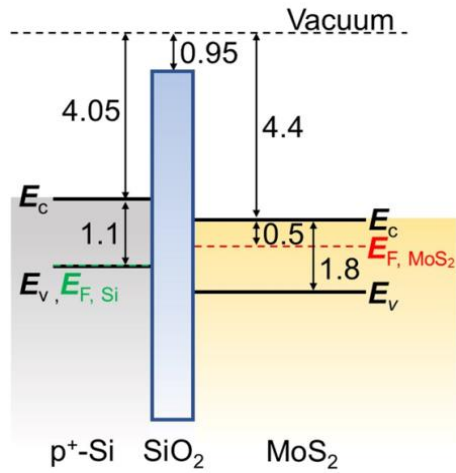

**Figure S17. Schematic of the band alignment of Si/SiO₂/MoS₂ based on KPFM measurement for device #5.**

**Supplementary Figure 18. STM topography of typical sulfur vacancies on MoS₂**

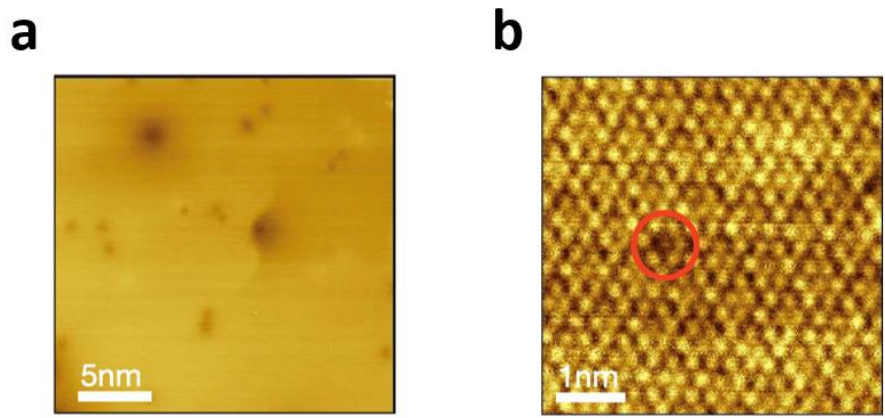

**Figure S18. STM topography of typical sulfur vacancies on MoS₂.** a) Surface topography (20 nm × 20 nm) with bias voltage −1 V, tunneling current 2.0 nA, and $B = 0$, where the defect density is $5.0 \times 10^{12}$ cm⁻². b) One instance of a sulfur vacancy marked by red circle observed under $B = 5$ T at 4.5 K with bias voltage of −0.4V and tunneling current of 2.0 nA.

**Supplementary Figure 19. Transport studies on h-BN buffered MoS₂ FETs**

To investigate the role of substrate and strain induced by the mismatching thermal expansion coefficient between $SiO_2$ substrate and $MoS_2$, we fabricated ML-MoS₂ FETs with a buffer layer of ~5 nm h-BN. The electrical transport results, as exemplified in Figure S19, shows that the h-BN buffered device did not exhibit any emergence of counterclockwise hysteresis under a magnetic field up to ±9 T, which differed fundamentally from all ML-MoS₂ FET devices directly transferred on to $SiO_2$/Si, while the universally existent clockwise hysteresis due to gate voltage stress was observed. These results suggest that with low lateral friction between MoS₂ and h-BN, lattice expansion of the buffered ML-MoS₂ trends to be isotropic such that no significant horizontal mirror symmetry breaking occurred under magnetic field-induced lattice expansion. Therefore, no flexoelectric effect induced ferroelectric-like behavior was seen.

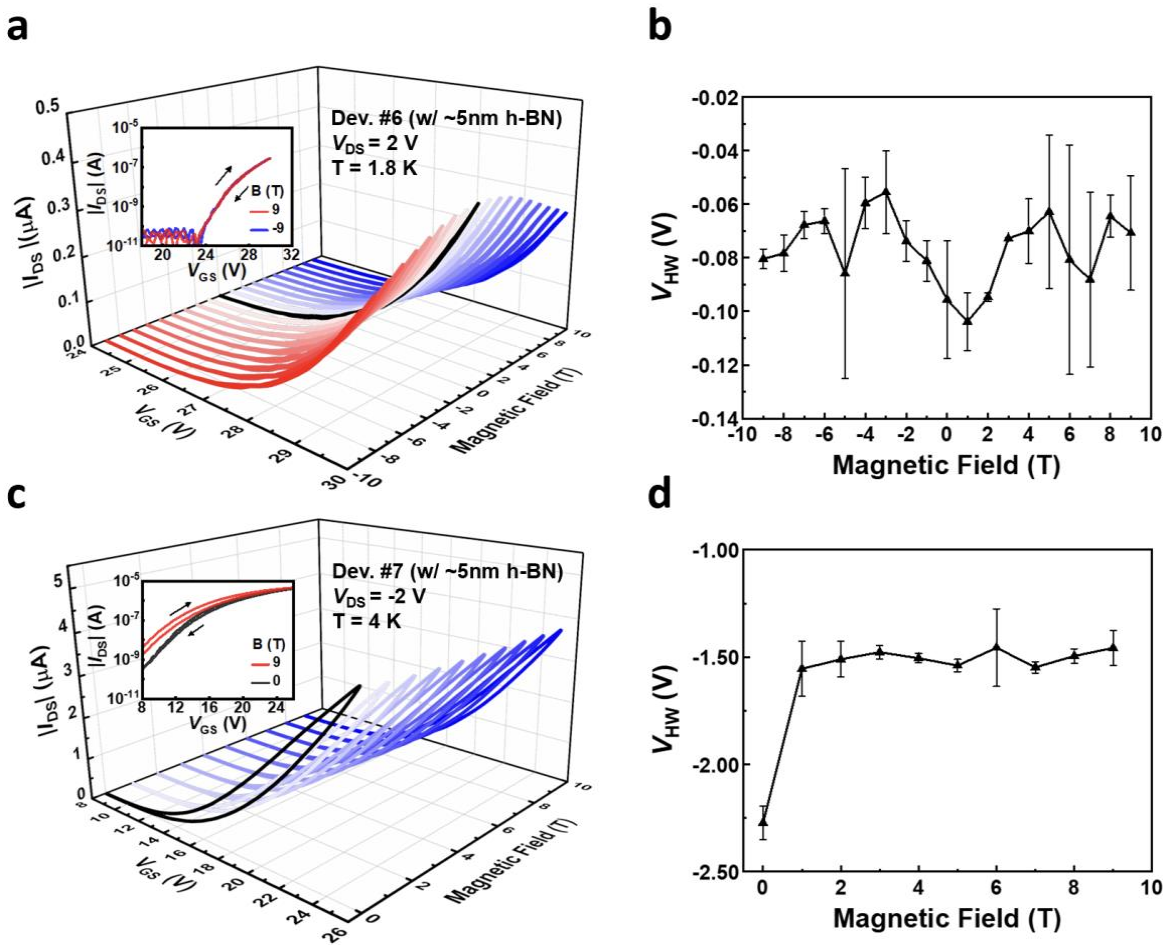

**Figure S19. Magnetic field-independent clockwise hysteresis at 1.8 K for h-BN buffered devices.** a) $|I_{DS}|$-*vs.*-$V_{GS}$ hysteresis under magnetic field from 9 T to –9 T with $V_{DS}$ = –1 V

measured at 1.8 K. As exemplified in the semi-log inset, clockwise hysteresis loop (indicated by black arrow) that was almost independent of magnetic field was observed under $B = 9$ and $B = -9$ T. b) $V_{HW}$-$vs.$-$B$ taken from a), showing that the negative value of $V_{HW}$ for clockwise hysteresis is independent of magnetic field. c) $|I_{DS}|$-$vs.$-$V_{GS}$ hysteresis under magnetic field from 0 T to $-9$ T with $V_{DS} = -2$ V measured at 4 K. As exemplified in the semi-log inset, clockwise hysteresis loop (indicated by black arrow) that was almost independent of magnetic field was observed under $B = 0$ and $B = -9$ T. d) $V_{HW}$-$vs.$-$B$ taken from c), showing that the negative value of $V_{HW}$ is independent of magnetic field except for the first few starting cycles at $B = 0$.